\documentclass[a4paper,leqno,12pt]{article}
\usepackage{a4wide}
\usepackage[utf8]{inputenc}
\usepackage{caption}
\usepackage{float, placeins, afterpage, rotating, graphicx, subcaption}
\usepackage{longtable, booktabs, tabularx, threeparttable}
\usepackage{svn-multi}
\usepackage{eurosym, calc}
\usepackage{amsmath, amssymb, amsfonts, amsthm, bm, delarray}
\usepackage{mathtools}
\usepackage{tikz}
\usepackage{pdfpages}
\usepackage{enumitem}
\usepackage{clipboard}
\usepackage{pdflscape}
\usepackage{multirow,array}
\usepackage[singlespacing]{setspace}
\usepackage{multirow}
\usepackage{bigdelim}
\usepackage{MnSymbol}
\usepackage[toc]{appendix}
\usepackage{breakurl}
\usepackage{mdframed}
\usepackage{natbib}
\usepackage[unicode=true]{hyperref}
\hypersetup{colorlinks=true, linkcolor=black, anchorcolor=black, citecolor=blue, filecolor=black, menucolor=black, runcolor=black, urlcolor=black}
\usepackage{setspace}
   
\usepackage[bottom]{footmisc}
\usepackage[left=2.5cm,top=2.5cm,bottom=2.5cm,right=2.5cm]{geometry}
\newcolumntype{L}[1]{>{\raggedright\arraybackslash}p{#1}}
\newcolumntype{C}[1]{>{\centering\arraybackslash}p{#1}}
\newcolumntype{R}[1]{>{\raggedleft\arraybackslash}p{#1}}

\newtheorem{hypothesis}{Hypothesis}
\makeatletter
\newcounter{subhyp} 
\let\savedc@hypothesis\c@hypothesis
\newenvironment{subhyp}
{%
	\setcounter{subhyp}{0}%
	\stepcounter{hypothesis}%
	\edef\saved@hypothesis{\thehypothesis}
	\let\c@hypothesis\c@subhyp     
	\renewcommand{\thehypothesis}{\saved@hypothesis\alph{hypothesis}}%
}
{}
\newcommand{\normhypothesis}{%
	\let\c@hypothesis\savedc@hypothesis 
	\renewcommand\thehypothesis{\arabic{hypothesis}}%
}

\makeatother
\begin{document}

\font\myfont=cmr12 at 12pt

\defcitealias{oecd2002leniency}{OECD, 2012}

\title{
Communication in the Infinitely Repeated Prisoner's Dilemma: Theory and Experiments\thanks{\protect\linespread{1}\protect\selectfont I thank Lisa Bruttel, Marco Caliendo, Sebastian Fehrler, Juri Nithammer, Vasilisa Petrishcheva, Tobias Werner as well as participants at the Economic Seminar at the University of Potsdam for very helpful comments and suggestions. Birte Prado Brand provided excellent research support. This research is funded by the Deutsche Forschungsgemeinschaft (DFG, German Research Foundation) -- project number 389447910, which I gratefully acknowledge. This project was preregistered at OSF before the data collection had started, see \nolinkurl{https://osf.io/xtmwv/?view_only=81a71835f0ca44019f4897a059825409}. }
}

\author{Maximilian Andres\thanks{\protect\linespread{1}\protect\selectfont Universität Potsdam. Email: maximilian.andres@uni-potsdam.de}}
\date{ 
	\today
}
\maketitle

\begin{abstract}

\singlespacing
So far, the theory of equilibrium selection in the infinitely repeated prisoner's dilemma is insensitive to communication possibilities.
To address this issue, we incorporate the assumption that communication reduces---but does not entirely eliminate---an agent's uncertainty that the other agent follows a cooperative strategy into the theory. 
Because of this, agents still worry about the payoff from cooperating when the other one defects, i.e.~the sucker's payoff $S$, and, games with communication are more conducive to cooperation than games without communication.
This theory is supported by data from laboratory experiments, and by machine learning based evaluation of the communication content.

\noindent
	\\
	\smallskip\\
	\textit{JEL-codes:} C73, C92, D83
\end{abstract}

\singlespacing

\clearpage
\doublespacing

\clearpage
\doublespacing

\section{Introduction}\label{sec:introduction}

Infinitely repeated prisoner’s dilemma games are useful to study cooperation in social dilemmas.\footnote{This game has been used to investigate cooperation, for example, in oligopolies \citep[see][]{mailath2006quantitysetting}, in environmental policy \citep[see][]{harstad2022}, in international relations \citep[see][]{powell1993internationalpolitics}, and in interpersonal relationships \citep[see][]{wilson2017example}.
}
Theoretical studies examine changes in the critical discount factor $\delta^{pd}$ to determine institutions that are more or less conducive to cooperation \citep[see][]{blonski2011rd}.

Experimental studies, however, suggest that this yields misleading results \citep[see][and the literature therein]{dalbo2018metainfinite}.
This is because the critical discount factor $\delta^{pd}$ does not account for the payoff from cooperating when the other one defects, i.e.~the sucker's payoff $S$.
The critical discount factor $\delta^{pd}$ is based on the assumption that agents are certain that the other agent follows a cooperative strategy. 
Hence, agents should ignore the payoff from cooperating when the other one defects ($S$).
\citet{blonski2011rd}, however, show that agents do consider the sucker's payoff $S$ by demonstrating that changes in the sucker's payoff $S$ do affect the level of cooperation in laboratory experiments.
The authors explain their finding by proposing that agents are maximally uncertain about the other agent's strategies and, hence, do account for the sucker's payoff~$S$.\footnote{
In a similar vein, \citet{dalbo2018metainfinite} point out that exclusively focusing on changes in the critical discount factor $\delta^{pd}$ yields inaccurate predictions for the level of cooperation in laboratory experiments.
In a meta study, the authors investigate the determinants of cooperation. 
For this aim, they include all the experimental papers they know of satisfying certain conditions. The stage game is a two by two prisoner's dilemma with perfect monitoring. The game is either one-shot or repeated using a random continuation rule which does not change during a supergame. Also the pairing of participants stays the same for a supergame. Additionally, the authors included only their own papers or studies published before 2014 if they had access to the data.  
The studies are: \citet{andreoni1993includedmeta, aoyagi2009infer,blonski2011rd,bruttel2012implement,cooper1996includedmeta,dalbo2005shadow,dalbo2010includedmeta,dalbo2011evolution,dalbo2019se,dreber2008includedmeta,duffy2009includedmeta,frechette2017includedmeta,fudenberg2012infer,kagel2013includedmeta,sherstyuk2013payment}.
Only recently, in multi-agent continuous time laboratory experiments, \citet{normann2021ninf} alternate the sucker's payoff $S$ to disentangle the effects of strategic uncertainty and strategic incentives in prisoner's dilemma games.
The authors find that changes in the sucker's payoff $S$ do affect cooperation, however, only in late supergames.
}

Yet, current literature neither aims to replicate their study nor looks at the interaction of this effect with communication: even though this interaction might lead agents to ignore the sucker's payoff $S$.
Experimental studies indicate that communication fosters the certainty of subjects about the cooperation of the other subjects \citep[see][and the literature therein]{dovrak2020comm}.
Incorporating this fact into a model suggests that agents will worry about the sucker's payoff $S$ as long as communication reduces, not eliminates, their uncertainty about the cooperative strategy of the other agent.
If, however, communication eliminates the uncertainty, the model suggests that agents disregard the sucker's payoff $S$.
Thus, according to the model, the effect of the sucker's payoff $S$ on cooperation depends on the effect of communication on uncertainty. 
This is why it is important to study both in conjunction.\footnote{Similar in spirit, \citet{dalbo2018metainfinite} conclude that it is of great interest to clarify the sufficiency of only looking at changes in the critical discount factor $\delta^{pd}$ to determine the level of cooperation once we allow communication.  
}

This paper studies, in laboratory experiments, the effect of changes in the sucker's payoff $S$ on cooperation in infinitely repeated prisoner’s dilemma games, both with and without communication.
For this aim, the laboratory experiments incorporate a two by two design that varies the sucker's payoff $S$ and whether subjects can communicate.
Comparing treatments with a high and a low sucker's payoff $S$ in games without communication may first replicate the results of \citet{blonski2011rd}. 
Additionally, to test whether this effect persists once we allow for communication, we compare treatments with a high and a low sucker's payoff $S$ in games with communication.
Finally, eliciting subject's beliefs and analyzing the content of the communication might shed light on the effect of communication on cooperation. 

Our results show that changes in the sucker's payoff $S$ affect cooperation in infinitely repeated prisoner's dilemma games, both with and without communication.
First, in line with \citet{blonski2011rd}, an increase in the sucker's payoff $S$ increases the level of cooperation in games without communication. 
Second, with communication, an increase in the sucker's payoff $S$ also increases the level of cooperation, but at an overall higher level of cooperation. 
Similarly to the literature \citep[see, e.g.,][]{blonski2011rd, dalbo2018metainfinite}, we conclude that theoretical studies that use infinitely repeated prisoner's dilemma games to determine the conduciveness of institutions to cooperation and disregard the sucker's payoff $S$, by only examining changes in the critical discount factor $\delta^{pd}$, may yield misleading results in settings, both, with and without communication.\footnote{In contrast to previous laboratory experiments, this paper studies the interaction of the role of the sucker's payoff $S$ and the beliefs with communication. Previous laboratory experiments focus on how much communication is needed to sustain cooperation \citep[see][]{dovrak2020comm} and the cooperation enhancing effect of communication \citep[see][]{kartal2021comm}. Note, in contrast to our paper,  \cite{kartal2021comm} refer to a model of endogenous preferences to explain the cooperation enhancing effect of communication. 
}

Explaining the results, we refer to our model.
In our model, changes in the sucker's payoff $S$ affect cooperation in games without communication, and in games with communication, but at an overall higher level of cooperation. This is because, according to our model, communication reduces, but does not eliminate, strategic uncertainty.\footnote{This explanation ties into a growing body of literature \citep[see, e.g.][]{aoyagi2020beliefs, normann2021ninf}, that suggests strategic uncertainty as a prominent reason for cooperation.}
That is in line with the data in two ways.
First, using an unsupervised machine learning algorithm to analyze the communication content, we provide suggestive evidence that subjects are indeed uncertain about the cooperation of the other subject before choosing an action.
Second, the analysis of beliefs suggests that communication reduces the uncertainty of subjects about the cooperation of the other subjects.

The remainder of this paper is organized as follows. In Section \ref{sec:theory}, we present the theory. In Section \ref{sec:design}, we describe the experimental design, and in Section \ref{sec:hypotheses}, we develop the hypotheses. In Section \ref{sec:results}, we present the results before we conclude in Section \ref{sec:conclusion}. An Appendix complements the article with the theoretical background (\ref{sec:model}), the instructions (\ref{sec:instructions}), the comprehension questions (\ref{sec:quiz}), the data (\ref{sec:additionalresults}) and the original tokens in their corresponding figures (\ref{sec:figures_in_german_kmeans}).

\section{Theory}\label{sec:theory}

This section presents the model. Before turning to the model, we formalize the stage game and the repetition procedure.

\paragraph{Stage game}

In a standard prisoner's dilemma $\Gamma$, two agents $i\in\{X,Y\}$ simultaneously face a choice $a$ between cooperation ($C$) and defection ($D$), $a_i\in\{C,D\}$. If both agents cooperate, each agent earns a reward payoff $R$. If both agents defect, each agent earns a punishment payoff $P$. If one agent defects while the other one cooperates, the defector earns a temptation payoff $T$ and the cooperator a sucker's payoff $S$. The stage game payoffs in the prisoner's dilemma $\Gamma$ are shown in Table \ref{table:stagegametheoryorginal}.

\begin{table}[H]
\centering
\setlength{\extrarowheight}{2pt}
\hspace*{\fill}
\subfloat[Original\label{table:stagegametheoryorginal}]{
\begin{tabular}{ccc}
\hline
                       & C                      & D                      \\ \cline{2-3} 
\multicolumn{1}{c|}{C} & \multicolumn{1}{c|}{R} & \multicolumn{1}{c|}{S} \\ \cline{2-3} 
\multicolumn{1}{c|}{D} & \multicolumn{1}{c|}{T} & \multicolumn{1}{c|}{P} \\ \cline{2-3} 
                       &                        &                        \\ \hline
\end{tabular}} \hspace*{\fill}  
\subfloat[Normalized\label{table:stagegametheorynormal}]{
\begin{tabular}{ccc}
\hline
                       & C                      & D                      \\ \cline{2-3} 
\multicolumn{1}{c|}{C} & \multicolumn{1}{c|}{$\frac{R-P}{R-P}=1$} & \multicolumn{1}{c|}{$\frac{S-P}{R-P}=-l$} \\ \cline{2-3} 
\multicolumn{1}{c|}{D} & \multicolumn{1}{c|}{$\frac{T-P}{R-P}=1+g$} & \multicolumn{1}{c|}{$\frac{P-P}{R-P}=0$} \\ \cline{2-3} 
                       &                        &                        \\ \hline
\end{tabular}}\hspace*{\fill}

    \caption{Row agent stage game payoffs in the prisoner's dilemma $\Gamma$.}
\label{table:stagegametheory}
\end{table}

Following \citet{rapoport1965payoffinequality}, a prisoner's dilemma $\Gamma$ features the conditions $T > R > P > S$ and $2 \cdot R > T+S$. The condition $T > R > P > S$ ensures that agents earn more from mutual cooperation than from mutual defection ($R>P$). It also guarantees that cooperation entails a risk to earn less ($P>S$) as each agent has an incentive to defect if the other agent chooses to cooperate ($T>R$). The condition $2 \cdot R > T+S$ ensures that mutual cooperation is more efficient than the asymmetric outcome.

A normalization of payoffs reduces the prisoner's dilemma $\Gamma$ to two parameters: gain $g$ from defection when the other agent cooperates and loss $l$ from cooperation when the other agent defects. The normalization of payoffs is accomplished by applying a monotonic linear transformation to the original payoffs. See \citet{stahl1991infeqpayoffs} and Table \ref{table:stagegametheorynormal} for a formalization.

\paragraph{Repetition}
An infinitely repeated prisoner's dilemma with discount factor $\delta$ can be seen as an interaction which continues with some probability. The probability that an interaction will continue is given by the discount factor $\delta$: $1 > \delta > 0$. Thus, the expected number of rounds equals $\frac{1}{1-\delta}$ \citep[see][]{roth1978first}.

Following \citet{dalbo2018metainfinite} and the literature therein, we assume that agents choose between two strategies at the inception of the game: grim and always defect.\footnote{Always defect is a plan of action or strategy which starts by defecting and continues to defect irrespective of the other agent's action. Grim starts by acting cooperatively and continues to act cooperatively until there is defection, then, defect forever. The latter is the strongest possible retaliation for defection by others. 
}
There is ample experimental evidence that a substantial fraction of subjects focus on grim and always defect \citep[see, e.g.,][]{dalbo2019se}.\footnote{The experimental evidence shows that the majority of subjects choose three pure-strategies: grim, always defect, and tit-for-tat. 
In games equal to and lower than $\delta = 0.75$, a majority of subjects choose grim and always defect \citep[see, e.g.,][]{dalbo2019se}. As we employ a continuation probability of $\delta = 0.75$ in the laboratory experiments, focusing on grim and always defect seems to be plausible.}
Following grim yields a value of $\frac{1}{1-\delta}$ when the other agent is following grim as well, and a value of $-l$ when the other agent is following always defect.
Whereas, following always defect yields a value of $1+g$ when the other agent is following grim as well, and a value of $0$ when the other agent is following always defect.
Hence, following always defect yields a value of $p \cdot  (1+g)$ and following grim yields a value of $p \cdot  (\frac{1}{1-\delta})+ (1-p) \cdot (-l)$.
$p$ \Copy{pexplenation}{is an agent's belief that the other agent is following} grim: $1 \geq p \geq 0$. Thus, $1-p$ \Paste{pexplenation} always defect.
Accordingly, the condition for which grim will be chosen is calculated as follows.
\begin{equation}\label{eqn:inequality1}
    p \cdot  \bigg(\frac{1}{1-\delta}\bigg)+ (1-p) \cdot (-l) 
    \geq  p \cdot  (1+g) 
\end{equation}
This condition holds if $\delta$ is larger than or equal to the critical discount factor $\delta^*$:
\begin{equation}\label{eqn:inequality3}
    \delta^*(g,l,p) = \frac{p \cdot (g-l)+l}{p\cdot(1+g-l)+l}
\end{equation}
Thus, the critical discount factor determines the minimum discount factor for which grim will be chosen.\footnote{Equation \ref{eqn:inequality3} shows the critical discount factor based on the idea of the size of the basin of attraction of always defect against grim by \cite{dalbo2018metainfinite}.}
Theoretical studies built on changes in the critical discount factor to predict cooperation (see, e.g., \citeauthor{athey2004collusionandprice}, \citeyear{athey2004collusionandprice}, \citeauthor{gilo2006partialcross}, \citeyear{gilo2006partialcross}).   
In this literature, a common interpretation of the critical discount factor is that cooperation is less likely reached and sustained when the value increases, and more likely when it decreases \citep[see][and the literature therein]{blonski2011rd}.

For any infinitely repeated prisoner's dilemma, we can calculate the minimum critical discount factor required to support cooperation in a sub-game perfect equilibrium in addition to defection by plugging $p=1$ into Equation \ref{eqn:inequality3}. The result is shown in Equation \ref{eqn:spe}. Hence, given that $\delta \geq \delta^{pd}$, there is a multiplicity of equilbria. This fact raises the question when agents will select the cooperative one.\newline

Following \citet{blonski2011rd}, this paper turns to investigating the equilibrium selection criteria Pareto dominance and risk dominance. Pareto dominance and risk dominance are equilibrium selection criteria to predict when agents will choose grim.

\paragraph{Pareto dominance} 
Grim is Pareto dominant if an agent is sufficiently patient given that the other agent follows grim: $p=1$. The critical discount factor $\delta^{pd}$, which is required for cooperation being a Pareto dominant response, can be calculated by plugging $p=1$ into Equation \ref{eqn:inequality3}. The result is shown in Equation \ref{eqn:spe}. 
\begin{equation}\label{eqn:spe}
\delta^{pd}(g) = \frac{g}{1+g}
\end{equation}

The condition for cooperation to be Pareto dominant is given by $\delta \geq \delta^{pd} $.

\paragraph{Risk dominance}

The grim strategy is risk dominant if it is a best response to the other agent, who is randomizing 50–50 between always defect and grim: $p=0.5$. The critical discount factor $\delta^{rd}$ required to support cooperation as a risk dominant response can be calculated by plugging $p=0.5$ into Equation \ref{eqn:inequality3}. The result is shown in Equation \ref{eqn:rd}. 
\begin{equation}\label{eqn:rd}
\delta^{rd}(g,l) = \frac{g+l}{1+g+l}
\end{equation}

For cooperation to be risk dominant, the condition $\delta \geq \delta^{rd}$ must be satisfied. Note, the risk dominant critical discount factor $\delta^{rd}$ is strictly larger than the Pareto dominant critical discount factor $\delta^{pd}$ \citep[see][]{blonski2011rd}.\newline

From a comparison of Equation \ref{eqn:spe} and Equation \ref{eqn:rd}, we can see that changes in the sucker's payoff $S$, i.e. the loss $l$, do affect risk dominance $\delta^{rd}$, but do not affect Pareto dominance $\delta^{pd}$. 
If $\overline{l} > \underline{l}$, then $\delta^{pd}$ in game $\Gamma(g,\underline{l})$ is equal to $\delta^{pd}$ in game $\Gamma(g,\overline{l})$. If $\overline{g} > \underline{g}$, then $\delta^{pd}$ in game $\Gamma(\underline{g},l)$ is lower than $\delta^{pd}$ in game $\Gamma(\overline{g},l)$. 
Risk dominance $\delta^{rd}$ in game $\Gamma(g,\underline{l})$ is lower than the risk dominance $\delta^{rd}$ in game $\Gamma(g,\overline{l})$. Similarly, risk dominance $\delta^{rd}$ in game $\Gamma(\underline{g},l)$ is lower than the risk dominance $\delta^{rd}$ in game $\Gamma(\overline{g},l)$.  
Thus, Equation \ref{eqn:spe} and Equation \ref{eqn:rd} 
show that a variation of $l$ affects risk dominance $\delta^{rd}$ and \textit{not} Pareto dominance $\delta^{pd}$, while a variation of $g$ affects both.

Note that both equilibrium selection criteria consider the same probability $p$ for games where agents can and cannot communicate. 
However, there is experimental evidence that communication results in an increase in the probability $p$ \citep[see][and the literature therein]{ellingsen2018oneshotcomm}.
One may question how this increase in the probability $p$ relates to the effect of changes in the sucker's payoff $S$, i.e.~the loss $l$, on the critical discount factor. 

\paragraph{Model}
To answer this question, we formally incorporate this effect into a model. 
First, we assume that risk dominance is useful in games without communication: $p=0.5$. There is ample experimental evidence suggesting that this criterion fares well in  determining the level of cooperation in games without communication \citep[see][and the literature therein]{dalbo2018metainfinite}.
Second, we assume that communication results in an increased belief $p^{+}$ for the other agent to follow grim (determined by the agent him- or herself): $1 > p^{+} > 0.5$. $p^{+}$ \Paste{pexplenation} grim in games with communication. Hence, $1-p^{+}$ \Paste{pexplenation} always defect in games with communication.
Thus, we incorporate into our model that the belief in games with communication is larger than in ones without communication: $p^{+} > 0.5$. It is important to note that the assumption entails that communication does \textit{not} cause an agent to be certain about the other agent following grim: $1 > p^{+}$.
Incorporating $p^{+}$ into Equation \ref{eqn:inequality3} results in Equation \ref{eqn:plus}.
\begin{equation}\label{eqn:plus}
    \delta^+(g,l,p^+) = \frac{p^+ \cdot (g-l)+l}{p^+\cdot(1+g-l)+l}
\end{equation}

Before investigating the effect of an increase in beliefs on the critical discount factor, we investigate whether a change in the sucker's payoff $S$, i.e. the loss $l$, affects the critical discount factor in games with communication.  
We find that a change in the loss $l$ ($\overline{l} > \underline{l}$) affects the critical discount factor in games with communication 
\begin{equation}\label{eqn:inequalitysuckerspayoff1}
    \delta^{+}_{\overline{l}} > \delta^{+}_{\underline{l}} 
\end{equation}
if and only if $1>p^{+}$. See Appendix \ref{sec:model_s} for a proof. $\delta^{+}$ in game $\Gamma(g,\underline{l})$ is shown as $\delta^{+}_{\underline{l}}$ and $\delta^{+}$ in $\Gamma(g,\overline{l})$ is shown as $\delta^{+}_{\overline{l}}$.\footnote{Following Table \ref{table:stagegametheoryorginal}, a higher sucker's payoff $S$ results in a lower $l$. Thus, $\overline{S}$ ($\underline{S}$) results in $\underline{l}$ ($\overline{l}$) for $\overline{S} > \underline{S}$.} 
Equation \ref{eqn:inequalitysuckerspayoff1} shows that the critical discount factor $\delta^+$ in game $\Gamma(g,\overline{l})$ is higher than in game $\Gamma(g,\underline{l})$ if and only if $1>p^{+}$.
In our model, given the assumption above that $1>p^{+}$, changes in the sucker's payoff $S$ do affect the critical discount factor ($\delta^{+}_{\overline{l}} > \delta^{+}_{\underline{l}}$).
Intuitively, if agents are uncertain ($1>p^{+}$) that the other agent is following the cooperative strategy, she or he must worry about the payoff from cooperating when the other agent defects: $l$. If, however, the agent is certain ($p=1$) about the cooperative strategy of the other agent, she or he must not consider the loss $l$ (see Equation \ref{eqn:spe} and the Pareto dominance equilibrium selection criterion).
Thus, according to our model, the effect of a change in the sucker's payoff $S$, i.e. the loss $l$, on the critical discount factor, depends on whether communication eliminates uncertainty or not.

Now, let us investigate the effect of an increase in beliefs on the critical discount factor. 
We find that the increase in beliefs by communication decreases the critical discount factor
\begin{equation}\label{eqn:inequalitycooperation1}
   \delta^{rd} > \delta^{+}
\end{equation}
if and only if $p^{+}>\frac{1}{2}$.
For a proof of Equation \ref{eqn:inequalitycooperation1}, see Appendix \ref{sec:model_c}. Equation \ref{eqn:inequalitycooperation1} shows that the critical discount factor in games without communication $\delta^{rd}$ is higher than the one in games with communication $\delta^{+}$ if $p^{+}>\frac{1}{2}$. Given our assumptions above ($p^{+}>\frac{1}{2}$), this condition is clearly satisfied in our model.\footnote{Intuitively, $\delta^{+} > \delta^{pd}$ as long as $1>p$. Given our assumptions above ($1>p^{+}$), this condition is clearly satisfied in our model and, hence, the critical discount factor in games with communication $\delta^{+}$ is larger than the Pareto dominant critical discount factor $\delta^{pd}$.}  
From Equation \ref{eqn:inequalitycooperation1}, it is apparent that the increase in beliefs by allowing for communication causes the value from following grim to be higher than the value from following always defect. This implies that the critical discount factor in games with communication is lower than in games without communication.  

Overall, according to our model, changes in the sucker's payoff $S$ do affect the critical discount factor as long as communication reduces, but does not eliminate, agents' uncertainty about the other agent following the cooperative strategy. 
If, however, communication does eliminate uncertainty, the model suggests that changes in the sucker's payoff $S$ do not affect the critical discount factor.
It is worth mentioning that the reduction of uncertainty results in a lower critical discount factor.
Hence, whether changes in the sucker's payoff $S$ do affect the critical discount factor depends on the effect of communication on uncertainty.

\section{Experimental design}\label{sec:design}

In the laboratory experiments, we vary the sucker's payoff $S$ and whether subjects can communicate in a between-subject design. Across the treatments, subjects participate in a number of infinitely repeated prisoner's dilemma games. Subjects interact in a number of supergames to grasp the payoff structure. 
In each supergame, two subjects interact continuously (partner matching protocol). Between each supergame, subjects are rematched using a perfect stranger matching protocol. 
The perfect stranger matching protocol eliminates the chance that a subject is recognizing another subject, which he or she has met before, based on his or her style to communicate. Recognizing others may set up a chance of reputation building and, hence, may affect the choice to cooperate. 
The stage game and continuation probability is similar to \citet{blonski2011rd}, who report stable results in their third supergame.\footnote{To adopt the stage game and continuation probability of \citet{blonski2011rd}, has the distinct advantage that we know how many supergames are needed for subject's to grasp the payoff structure and, hence, allows us to choose the number of supergames accordingly. 
} Accordingly, there should be at least three supergames per treatment. We employ five supergames per treatment. Ergo, a perfect (stranger) matching graph involves six subjects \citep[see][and the literature therein]{both2016psm}. 

\paragraph{Communication} In each supergame, before subjects set their action $a_i$ in round one, a free-form chat window opens for $60$ seconds. A pre-play free-form chat window enables subjects to negotiate their strategy choice and, then, to choose a strategy for a supergame.\footnote{Communication in every round, for example, would enable agents to say sorry for unilateral defection and, hence, enable them to re-negotiate their strategy. To hinder them to re-negotiate, we choose pre-play communication \citep[see][]{farrell1989renegotiation}.
} 

\paragraph{Stage Game} After the free-form chat window closes, both subjects choose their action $a_i$ simultaneously in the infinitely repeated prisoner's dilemma $\Gamma(T=100,R=90,P=80,S)$.\footnote{Subjects choosing actions and not strategies has the distinct advantage that we can compare our results to those of \citet{blonski2011rd} and other studies. In the literature, it is common that subjects choose actions and not strategies. According to \citet{dalbo2011evolution} and \citet{dalbo2018metainfinite}, subjects focus on grim and always defect in those studies.
} To provide a neutral frame, $a_i=\{C,D\}$ is renamed into $a_i=\{A,B\}$.
Following each round, subjects receive feedback about their own action, the action of the other agent and their own payoff in that round and the supergame so far.

\paragraph{Repetition} After every round $r \geq 2$ the game continues with a probability $\delta$ and ends with a probability $1- \delta$. A probability $\delta^{'}$ is drawn from a uniform distribution over $[0, 1]$. If and only if $\delta^{'} \leq \delta$ another round starts for all pairs. 
The continuation probability is $\delta = 0.75$.

\paragraph{Treatments} We employ a between-subject design varying the sucker's payoff $S$ and whether agents can communicate.\footnote{To eliminate the chances of an experimenter demand effect, a between-subject design and not a within-subject design \citep[as in][]{blonski2011rd} is employed. Here, a within-subject design may result in a non-constant experimenter demand effect between treatments where people can and can \textit{not} communicate. The reason is that communication can be used to discuss what the experimenter wants and, hence, it may be more salient in treatments where people can communicate than in treatments where they can not. 
} Thus, we consider four treatments: \textsc{NoComm70}, \textsc{NoComm0}, \textsc{Comm70} and \textsc{Comm0}. In \textsc{NoComm70} no free-form chat window opens and the sucker's payoff equals $70$ and in \textsc{NoComm0}, no free-form chat window opens and the sucker's payoff equals $0$. In \textsc{Comm70} a free-form chat window opens and the sucker's payoff equals $70$ while in \textsc{Comm0} a free-form chat window opens too, but the sucker's payoff equals $0$. 
Table \ref{table:experimentdesign} summarizes the experimental design.

\begin{table}[H]
\begin{center}
\footnotesize
\begin{tabular}{cccccc}
\hline
                                                                                   &                                    & \multicolumn{4}{c}{\multirow{2}{*}{Treatment}}                                                          \\
                                                                                   &                                    & \multicolumn{4}{c}{}                                                                                    \\ \cline{3-6} 
                                                                                   &                                    & \multirow{2}{*}{\textsc{NoComm70}} & \multirow{2}{*}{\textsc{NoComm0}} & \multirow{2}{*}{\textsc{Comm70}} & \multirow{2}{*}{\textsc{Comm0}} \\
                                                                                   &                                    &                           &                          &                         &                        \\ \hline
\multirow{4}{*}{\begin{tabular}[c]{@{}c@{}}Between-Subject\\ Design\end{tabular}}  & \multirow{2}{*}{Communication}     & \multirow{2}{*}{no}        & \multirow{2}{*}{no}      & \multirow{2}{*}{yes}      & \multirow{2}{*}{yes}     \\
                                                                                   &                                    &                           &                          &                         &                        \\
                                                                                   & \multirow{2}{*}{Sucker's payoff $S$} & \multirow{2}{*}{70}       & \multirow{2}{*}{0}       & \multirow{2}{*}{70}     & \multirow{2}{*}{0}     \\
                                                                                   &                                    &                           &                          &                         &                        \\ \hline
\multirow{4}{*}{\begin{tabular}[c]{@{}c@{}}Theoretical \\ Prediction\end{tabular}} & \multirow{2}{*}{$\delta^{pd}$}               & \multirow{2}{*}{0.50}     & \multirow{2}{*}{0.50}    & \multirow{2}{*}{0.50}   & \multirow{2}{*}{0.50}  \\
                                                                                   &                                    &                           &                          &                         &                        \\
                                                                                   & \multirow{2}{*}{$\delta^{rd}$}                & \multirow{2}{*}{0.67}     & \multirow{2}{*}{0.90}    & \multirow{2}{*}{0.67}   & \multirow{2}{*}{0.90}  \\
                                                                                   &                                    &                           &                          &                         &                        \\

\hline
\end{tabular}
\caption{Between-subject design and theoretical predictions split up by treatments.}
\label{table:experimentdesign}
\end{center}
\end{table}
\paragraph{Probability}
Each subject $i$ is asked immediately after having chosen their action and before they receive feedback to state their probability $\Tilde{p}_i$ 
that the other agent will cooperate, on an intuitive slider without a default. 
A subject's belief about the others action in round one is a proxy for her or his assessment about the others strategy choice.\footnote{For example, if a subject is certain that the other subject cooperates in the first round, he or she is certain that the other subject is following a cooperative strategy such as grim.} 
For the following reasons, we do not incentives the belief elicitation procedure. 
First, due to the evidence that incentivized belief elicitation affects subsequent actions choices and the chance of this influencing the subsequent action choices differently for a change in the sucker's payoff $S$ and a communication possibility, respectively \citep[see][]{gachter2010beliefspgg}.
Second, due to the ample experimental evidence that communication can increase an agents belief about the cooperation of the other agent \citep[see][and the literature therein]{ellingsen2018oneshotcomm}. Third, to opt for simplicity whenever possible \citep[see][]{aoyagi2020beliefs}. 
Following \citet{gill2020beliefelicit}, to make the belief elicitation procedure as minimally invasive as possible, we only elicit beliefs in the first round of the first and the last supergame. 
This procedure aims to prevent any contamination on subsequent action choices by the probability elicitation stage.

\paragraph{Procedure} Assignment to different treatments is random in the sense that subjects signing up for a session do not know which treatment is run. Before the experiment starts, subjects are seated randomly at computer terminals. Instructions are given in written form. The instructions are presented in Appendix \ref{sec:instructions}. 
\Copy{quiz}{After the instructions are read, subjects are asked comprehension questions on the screen to ensure and to make it common knowledge they all understand the important parts of the experiment.} The comprehension questions are presented in Appendix \ref{sec:quiz}. Only after all subjects passed the comprehension questions, the experiment starts.
The experiment was conducted in May $2022$ at the University of Potsdam and a total of $132$
students participated. 
The subject's final earnings are the sum of their payoffs in points, plus a show-up fee. They earned (participated), on average, $14.83$ euro ($36$ minutes) with a minimum of $9.80$ euro ($29$ minutes) and a maximum of $22.00$ euro ($45$ minutes). Across the subjects, $30$ participated in \textsc{NoComm70}, \textsc{Comm70} and \textsc{Comm0}, respectively, and $42$ in \textsc{NoComm0}. 
Similar to \citet{blonski2011rd}, in total, we observe 1.401 stage game interactions.

\section{Hypotheses}\label{sec:hypotheses}

In the following, we set up hypotheses to state the effect of changes in the sucker's payoff $S$ and the effect of communication on the rate of cooperation, respectively. 
Before turning to these hypotheses, we introduce a set of hypotheses on the effect of communication on beliefs. 
In our model, we argue that communication fosters certainty: the belief in games with communication ($p^{+}$) is higher than in games without communication ($p=0.5$). Thus, we expect that beliefs in treatments with communication are, on average, higher than in ones without communication.  
\begin{subhyp}
\begin{hypothesis}\label{hypothesisa1}
The mean belief in \textsc{Comm70} is higher than in \textsc{NoComm70}.\footnote{This hypothesis was not preregistered.\label{footnote:hypothesis}}
\end{hypothesis}
\begin{hypothesis}\label{hypothesisa2}
The mean belief in \textsc{Comm0} is higher than in \textsc{NoComm0}.$^{\ref{footnote:hypothesis}}$
\end{hypothesis}
\end{subhyp}
Next, this section introduces the hypotheses on the effect of changes in the sucker's payoff $S$, i.e., the loss $l$, on the rate of cooperation. 
It is apparent from Table \ref{table:experimentdesign} that changes in the loss $l$, do not affect the Pareto dominance critical discount factor $\delta^{pd}$. 
While according to our model, changes in the loss $l$ do affect the critical discount factor in games with communication ($\delta^{+}$) due to $1 > p^{+}$ and in games without communication ($\delta^{rd}$). 
Thus, according to our model, we expect the rate of cooperation in treatments with a sucker's payoff equal to $70$ to be higher than in treatments with a sucker's payoff equal to $0$ in games with communication and in ones without communication, respectively.

\begin{subhyp}
\begin{hypothesis}\label{hypothesisb1}
The rate of cooperation in \textsc{Comm70} is higher than in \textsc{Comm0}.
\end{hypothesis}
\begin{hypothesis}\label{hypothesisb2}
The rate of cooperation in \textsc{NoComm70} is higher than in \textsc{NoComm0}.
\end{hypothesis}
\end{subhyp}
Finally, this section turns to the hypotheses on the effect of communication on the rate of cooperation. Table \ref{table:experimentdesign} shows that the Pareto dominance critical discount factor $\delta^{pd}$ in games with communication is equal to the one in games without communication. According to our model, however, the critical discount factor in games without communication is higher than in ones with communication ($\delta^{rd} > \delta^{+}$) in $\Gamma(g,\underline{l})$ and $\Gamma(g,\overline{l})$, respectively, for $\overline{l} > \underline{l}$. This aspect of the model indicates that the rate of cooperation in treatments with communication is higher than in ones without communication.    
\begin{subhyp}
\begin{hypothesis}\label{hypothesisc1}
The rate of cooperation in \textsc{Comm70} is higher than in \textsc{NoComm70}.$^{\ref{footnote:hypothesis}}$
\end{hypothesis}
\begin{hypothesis}\label{hypothesisc2}
The rate of cooperation in \textsc{Comm0} is higher than in \textsc{NoComm0}.$^{\ref{footnote:hypothesis}}$
\end{hypothesis}
\end{subhyp}

Following \citet{cooper2022communication} and \citet{kartal2021comm}, we analyze the communication content to better understand why subjects made certain choices. 
This analyzes may provide suggestive evidence on whether subjects are indeed uncertain about the cooperation of the other subject.

\section{Results}\label{sec:results}

In this section, we first study the effect of communication on beliefs. Second, we study the effect of changes in the sucker's payoff $S$ on cooperation in games with and without communication, respectively. We then continue to investigate the effect of communication on cooperation. Finally, this section turns to the analyses of the communication content.

\paragraph{Beliefs}

\begin{figure}[ht]
\begin{minipage}[c]{\textwidth}
		\begin{subfigure}[h]{0.49\textwidth}
        \includegraphics[trim=0  1cm 0 0, clip=true,width=\linewidth]{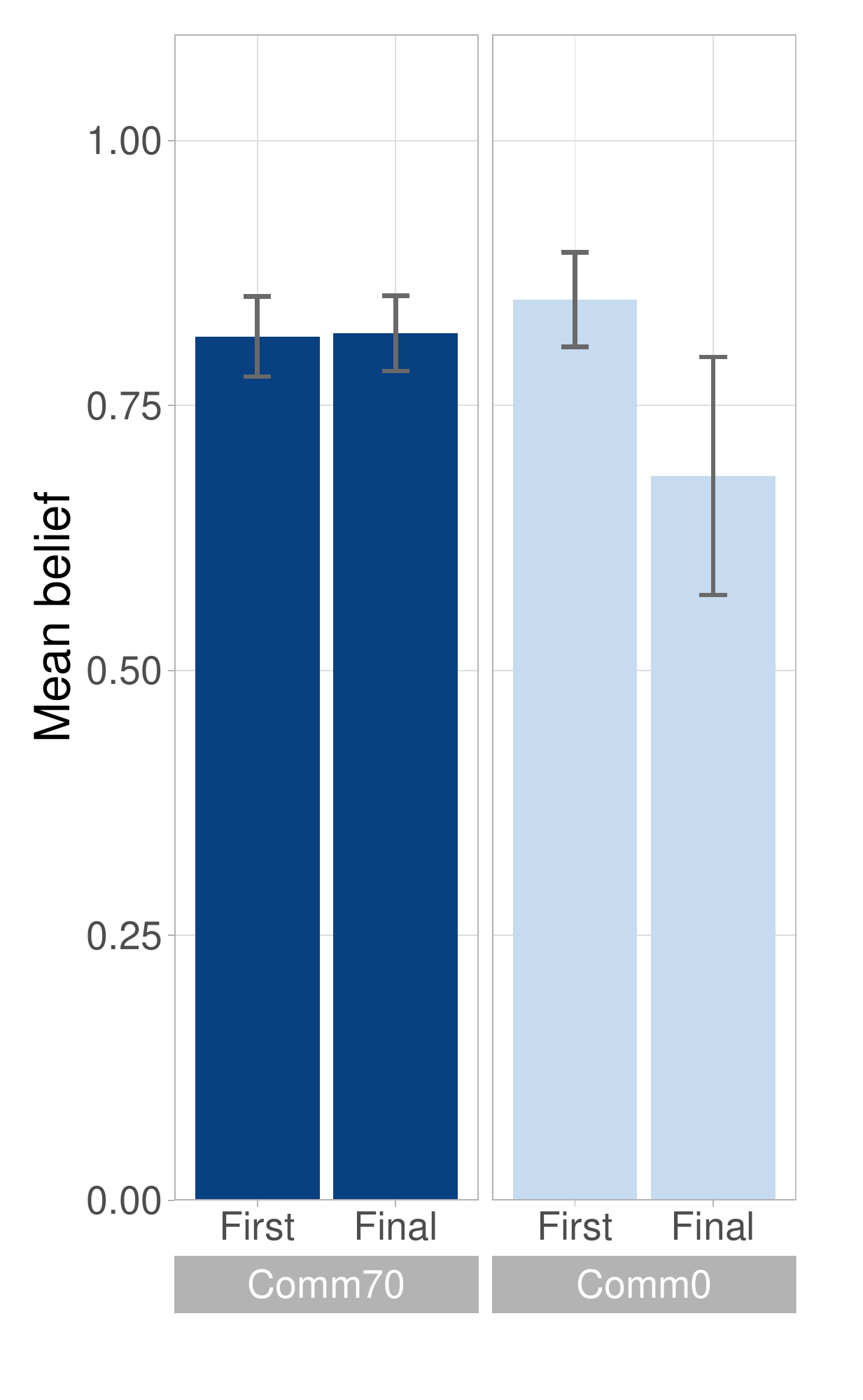}
        \caption{Communication}
			\label{fig:beliefcomm}
		\end{subfigure}	
	~
		\begin{subfigure}[h]{0.49\textwidth}
        \includegraphics[trim=0  1cm 0 0, clip=true,width=\linewidth]{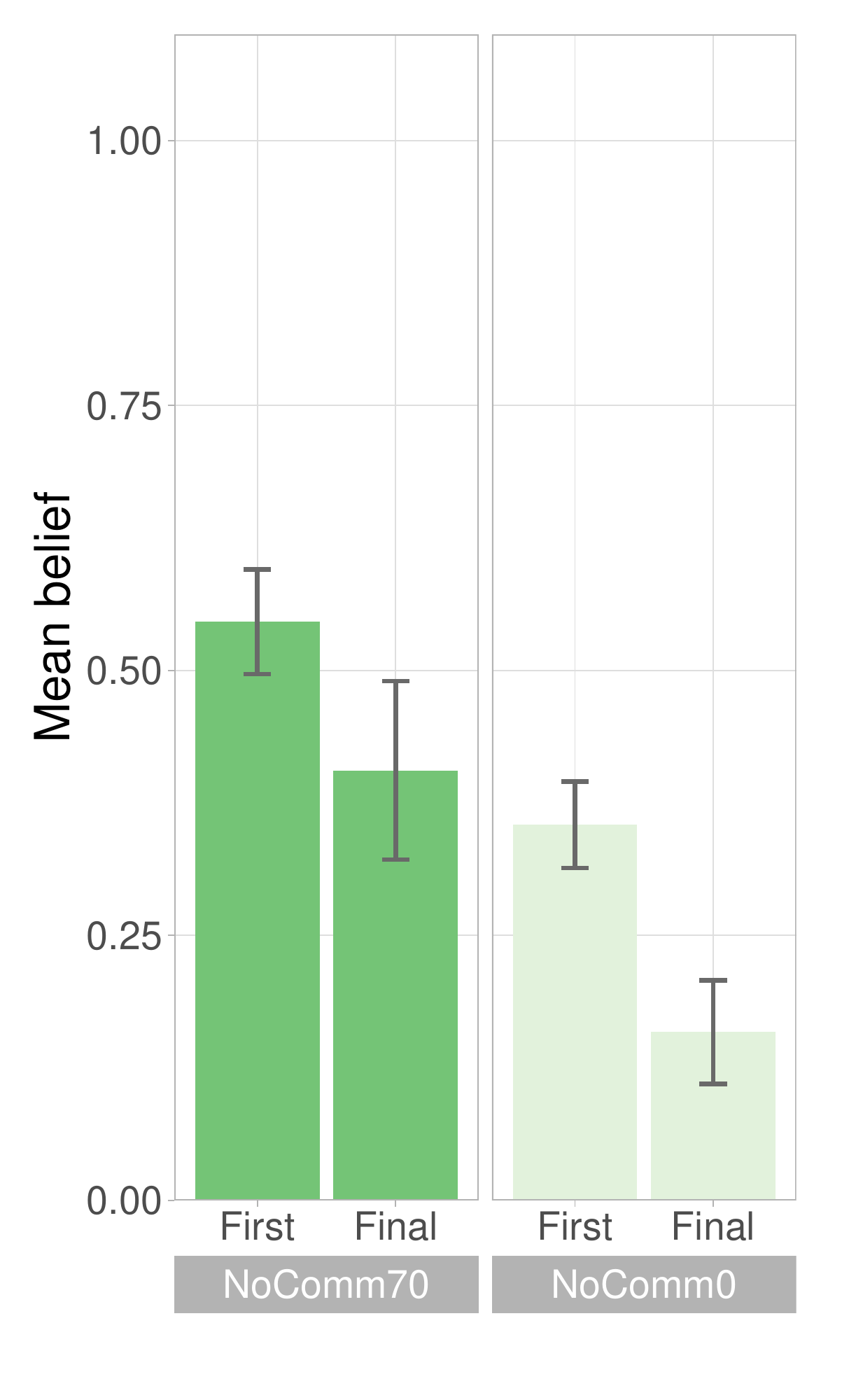}
        \caption{No communication}
			\label{fig:beliefnocomm}
		\end{subfigure}
\end{minipage}	
\caption[The \Paste{beliefigure}]{The \Paste{beliefigure} Error bars indicate standard errors, clustered at the graph level. Table  \ref{table:beliefmedian} in Appendix \ref{sec:additionalresults} provides further information.}\label{fig:beliefsupergames}
\end{figure}

Figure \ref{fig:beliefsupergames} presents the \Copy{beliefigure}{mean belief in the first and final supergame split up by treatments.} It is apparent from this figure that the mean belief in treatments with communication is substantially higher than in ones without communication. A one-sided Wilcoxon-Mann-Whitney test with continuity correction\footnote{Similar to \citet{blonski2011rd}, unless noted otherwise, all $p$-values reported in this paper refer to a one-sided Wilcoxon-Mann-Whitney test with continuity correction and graph-level clustering. The continuity correction accounts for the discontinuity in small sample sizes and produces more conservative $p$-values.} shows that the mean belief in the first supergame in \textsc{Comm70} (\textsc{Comm0}) is significantly higher than in \textsc{NoComm70} (\textsc{NoComm0}): $p<0.001$, $N=15$ ($p<0.001$, $N=15$ and $N=21$, respectively). The result is very similar if we instead consider the final supergame: $p=0.010$ ($p=0.005$). 
Thus, our data clearly supports Hypothesis \ref{hypothesisa1} and Hypothesis \ref{hypothesisa2} that the mean belief in treatments with communication is higher than in ones without communication.

Another observation is worth mentioning.
As can be seen from Figure \ref{fig:beliefcomm}, the mean belief in treatments with communication is below one and above one half, across supergames. 
This suggests that neither Pareto dominance, implies a belief level of one, nor risk dominance, implies a belief level of one half, is a useful equilibrium selection criterion in settings with communication.

Taken together, the fact that the mean belief is higher in treatments with communication than in treatments without communication and the suggestive evidence that the mean belief in treatments with communication is below the certainty level of one and above the level of maximum uncertainty of one half, indicates that communication reduces, not eliminates, uncertainty. 
Thus, our data is in line with the proposed effect of communication on uncertainty presented in Section \ref{sec:theory}.

\paragraph{Cooperation}\label{sec:cooperation} 

\begin{figure}[ht]
\begin{minipage}[c]{\textwidth}
		\begin{subfigure}[h]{0.49\textwidth}
        \includegraphics[trim=0  0cm 0 0, clip=true,width=\linewidth]{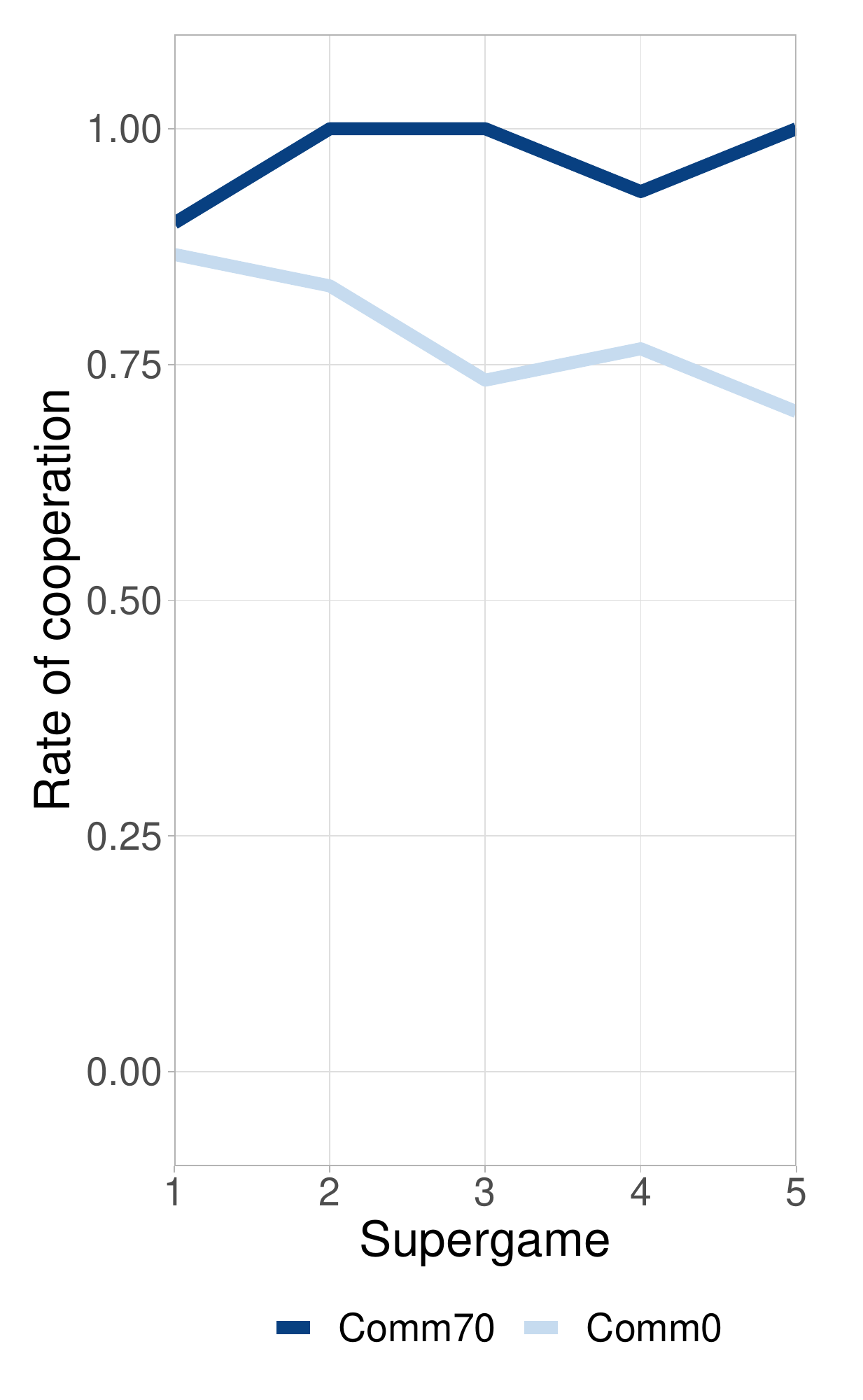}
        \caption{Communication}
			\label{fig:cooperationcomm}
		\end{subfigure}	
	~
		\begin{subfigure}[h]{0.49\textwidth}
        \includegraphics[trim=0  0cm 0 0, clip=true,width=\linewidth]{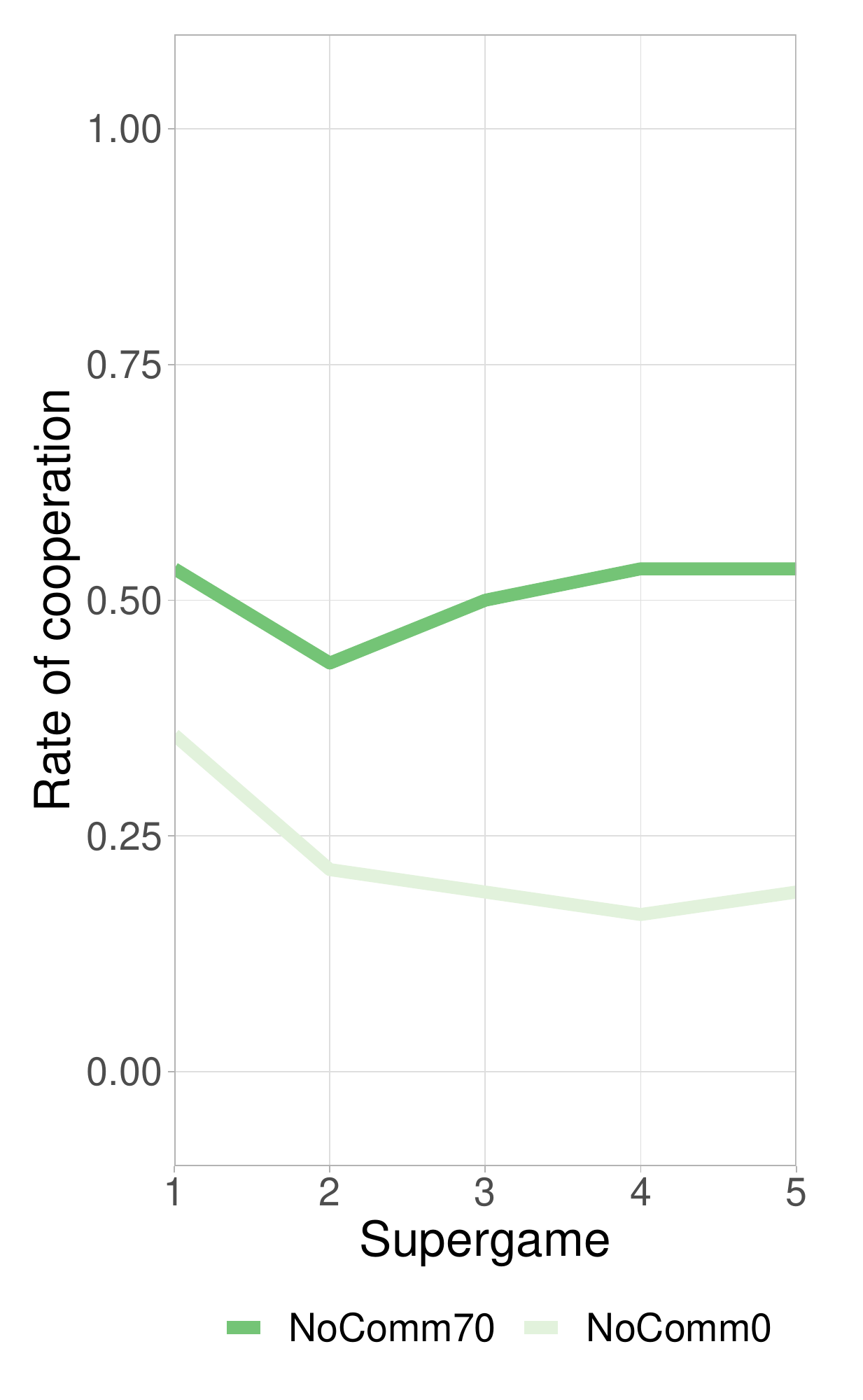}
        \caption{No communication}
			\label{fig:cooperationnocomm}
		\end{subfigure}
\end{minipage}	
\caption{The \Paste{cooperationsupergames}}\label{fig:cooperationsupergames}
\end{figure}

Figure \ref{fig:cooperationsupergames} shows the \Copy{cooperationsupergames}{rate of cooperation in round one over supergames, split up by treatments.} 
A subject's cooperation choice in the first round is a proxy for her or his strategy choice in that supergame.
From Figure \ref{fig:cooperationcomm}, we can see that the rate of cooperation in \textsc{Comm70} is higher than in \textsc{Comm0} in late supergames. A test shows that the rate of cooperation in round one in the final supergame in \textsc{Comm70} is significantly higher than in \textsc{Comm0}: $p=0.035$. The result is similar over all supergames: $p=0.055$. The rate of cooperation in round one in the first supergame, however, is not significantly higher in \textsc{Comm70} than in \textsc{Comm0}: $p=0.409$.\footnote{Also if we compare first round actions in supergame three (two), the rate of cooperation is (marginally) significantly higher in \textsc{NoComm70} than in \textsc{NoComm0}: $p=0.012$ ($p=0.090$). While the result in the fourth supergame in \textsc{Comm70} is higher than in \textsc{Comm0}, it is not statistically significant ($p=0.144$).} The result is very much the same if we instead look at all rounds. See Table \ref{table:cooperationsupergamesallrounds} in Appendix \ref{sec:additionalresults} for support.
Thus, in late supergames, our data supports Hypothesis \ref{hypothesisb1} that changes in the sucker's payoff $S$ affect the rate of cooperation in games with communication.

As can be seen from Figure \ref{fig:cooperationnocomm}, in round one, the rate of cooperation in \textsc{NoComm70} is higher than in \textsc{NoComm0} across supergames. We find that the rate of cooperation in the first round in the final supergame in \textsc{NoComm70} is significantly higher than in \textsc{NoComm0}: $p=0.032$. Checking all supergames ($p=0.025$), we find comparable results. In the first supergame, however, the rate of cooperation is not significantly ($p=0.375$) higher in \textsc{NoComm70} than in \textsc{NoComm0}.\footnote{A test shows that the rate of cooperation in round one in supergame three (four) in \textsc{NoComm70} is significantly higher than in \textsc{NoComm0}: $p=0.021$ ($p=0.035$). While this is not the case in the second supergame ($p=0.135$), the mean rate of cooperation points in the predicted direction.} The results are almost the same if we alternatively examine the rate of cooperation in all rounds. To back this up, see Table \ref{table:cooperationsupergamesallrounds} in Appendix \ref{sec:additionalresults}. Ergo, in late supergames, our data supports Hypothesis \ref{hypothesisb2} that changes in the sucker's payoff $S$ affect the rate of cooperation in games without communication.  

It is apparent from Figure \ref{fig:cooperationsupergames} that the rate of cooperation in treatments with communication is higher than in ones without communication across supergames.  
A test shows that the rate of cooperation in round one of the final supergame (over supergames) in \textsc{Comm70} is significantly higher than in \textsc{NoComm70}, $p=0.012$ ($p=0.006$). The result is very much the same if we rather consider the rate of cooperation in round one of the final supergame (over supergames) between \textsc{Comm0} and \textsc{NoComm0}: $p=0.015$ ($p=0.005$). Even if we look at round one of the first supergame, the rate of cooperation in \textsc{Comm70} (\textsc{Comm0}) is significantly higher than in \textsc{NoComm70} (\textsc{NoComm0}): $p=0.004$ ($p=0.003$).\footnote{It is apparent from this analyses, that the rate of cooperation in round one in \textsc{Comm70} (\textsc{Comm0}) is significantly higher than in \textsc{NoComm70} (\textsc{NoComm0}) in the second, third and fourth supergame: $p=0.003$, $p=0.004$ and $p=0.022$ ($p=0.004$, $p=0.003$ and $p=0.006$), respectively.} The result is almost identical if we instead consider the rate of cooperation in all rounds. See Table \ref{table:cooperationsupergamesallrounds} in Appendix \ref{sec:additionalresults} for support.
Thus, our data clearly supports Hypothesis \ref{hypothesisc1} and Hypothesis \ref{hypothesisc2} that the rate of cooperation in treatments with communication is higher than in ones without communication in games with a sucker's payoff of $70$ and a sucker's payoff of $0$, respectively.

\paragraph{Communication}\label{sec:communication}

Before studying the communication content, let us first explain our approach to analyze the text data. 
To investigate the communication content, we use an unsupervised machine learning algorithm. 
This algorithm has the distinct advantage that it does not rely on pre-defined clusters, which are typically introduced through a process of human hand-coding. Human hand-coding is the most common approach to cluster chat data in experimental economics \citep[see][]{andres2022ml}. 
Such a process involves defining clusters based on an analysis (of a subset) of the chat data, and then having raters assign the data to these pre-defined clusters. Characterizing clusters manually, however, may be subject to biases, for example, in the interpretation of the definition of clusters \citep[see, e.g.,][]{kartal2021comm}.
For this reason, in addition to human hand-coding, we use an unsupervised machine learning algorithm to specify clusters.  
 
Our approach starts by looking at the entire communication content across supergames and treatments. Thus, we obtain $150$ chats or documents, which together form the corpus. 
The corpus is subject to a systematic natural language procedure, which includes the correction of spelling mistakes, the reduction of words to their dictionary form, and the elimination of words that are not meaningful. 
The processed corpus can be represented in a matrix $\Lambda$, where the element $( \mu, \theta, \omega)$ shows the word embedding value $\omega$ of the unique token $\theta$ that appeared in the document $\mu$.
A token $\theta$ can be a word or, for example, or a number.
A word embedding of a token $\omega_{\theta}$ is a real-valued vector, which encodes the meaning of the token $\theta$ such that tokens that are closer in the vector space are expected to be similar in meaning \citep[see][]{joulin2016bag}.
The matrix $\Lambda$ is subject to the sum operator such that the element $( \mu, \omega)$ shows the sum of the word embedding $\omega$ for each token $\theta$ in each document $\mu$. 
This representation $\overline{\Lambda}$ has been proven particularly useful to
cluster documents with similar meaning \citep[see][and the literature therein]{ash2022wordembeddings}.

To cluster documents with similar meaning, we use $k$-means. This unsupervised machine learning algorithm is especially suitable for separating a corpus into $k$ clusters \citep[see][]{steinbach2000kmeans}. 
Documents within the same cluster are as similar as possible, whereas documents from different clusters are as dissimilar as possible.
The basic idea behind this algorithm is to define clusters such that the total within-cluster variation is minimized.
To identify the number of clusters $k$, we examine the total-within clusters sum of square. The results of this method suggest two clusters: 
See Figure \ref{fig:elbowcurve} in Appendix \ref{sec:additionalresults} for support. Hence, we use two clusters to investigate the communication content.

\begin{figure}[h]
	\centering
	\includegraphics[width=1\linewidth]{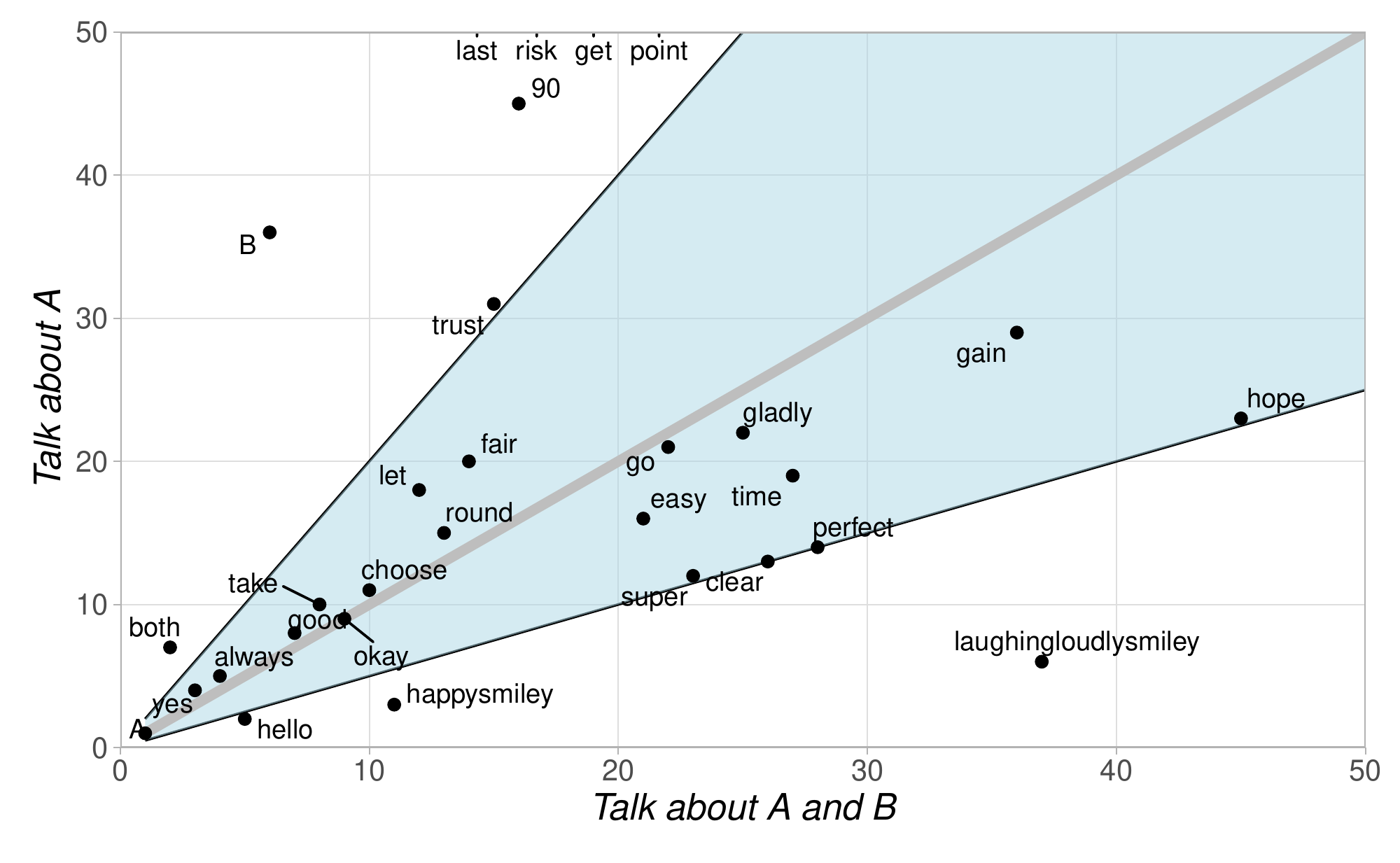}
	\caption{Frequency rankings of the 30 most used tokens in both clusters. Figure \ref{fig:figures_in_german_kmeans} in Appendix \ref{sec:figures_in_german_kmeans} provides the tokens in German.}
 \caption*{\textit{Note: The two black lines indicate the border, where the two relative rank differentials $\frac{r_{2}-r_{1}}{r_{1}}$ and $\frac{r_{1}-r_{2}}{r_{2}}$, $r_{1}$ and $r_{2}$ are the rank in cluster one and in cluster two, respectively, are equal to 1. Tokens outside the shaded area have a relative rank differential exceeding 1. Tokens are more frequent in one cluster than in the other if the relative rank statistic is larger than or equal to 1 \citep[see][]{fischer2019rrs}.}}
	\label{fig:rrds}
\end{figure}

In Figure \ref{fig:rrds}, we depict the $30$ most frequent tokens in the corpus and their relative rank differential to show the key distinguishing features between both clusters.
It is apparent from this figure, 
that the key distinguishing feature between the two clusters is 
whether subjects engage in a discussion about the riskiness of the action choices.
In the \textit{Talk about A and B} cluster, the tokens `B', `risk' and 'trust' often appear together with the token 'A', indicating that subjects mainly discuss the riskiness of action choices.
In the \textit{Talk about A} cluster, the token `A' often appears together with tokens related to agreeableness (`perfect', `super' and `happy-smiley'), indicating that subjects mainly talk about an agreement to choose the cooperative action.
The fact that words like `risk', `trust' and the defection action `B' are present in one cluster, but not in the other one, indicates that subjects are uncertain about the action of the other subject before choosing an action -- at least in one cluster.

Compared to our machine learning approach presented above, we find qualitatively similar results with human hand-coding.
The hand-coding was realized as follows. 
First, two student research assistants, independently from each other, read the corpus to note clusters present in the data. 
Then, they met and agreed on two relevant clusters: agreement and discussion. The first cluster---agreement---captures chats where the subjects purely agree to choose action `A', i.e. to cooperate. In the second cluster---discussion---subjects communicate on both actions, `A' and `B', and the dilemma-aspect of the game.  
Second, the student assistants independently clustered every chat to the respective clusters:
Their work across all clusters is consistent ($\kappa = 0.71$) with each other.
The Cohen's $\kappa$ across all clusters between the human raters and our machine learning approach is above $0.62$, indicating a substantial agreement. 
Thus, the data supports our machine learning approach, indicating that subjects are uncertain about the cooperation of the other subject before choosing an action.

To sum up, in line with the machine learning approach, a communication analysis by human hand-coding indicates that subjects are uncertain about the cooperation of the other subject before choosing an action. 
It is worth mentioning that we find qualitatively similar results between the human hand-coding and machine learning approach.
This finding provides further evidence in line with our assumption that subjects are uncertain about the cooperative strategy of the other subject.

\section{Conclusion}\label{sec:conclusion}

This paper studied, in laboratory experiments, whether incorporating the assumption that communication increases the belief, but does not causes certainty that the other agent cooperates, into the theory of equilibrium selection in the infinitely repeated prisoner's dilemma, yields a criterion $\delta^{+}$, which fares much better as a tool for predicting cooperation than traditional ones.
Traditional ones are insensitive to communication possibilities. 
Our results suggest that the equilibrium selection criterion $\delta^{+}$ is a better tool for predicting the effects of communication possibilities than traditional ones.  

We expect this paper to be useful in at least two ways: 
First, there is an influential body of literature on the predictability of cooperation to which we add here, using equilibrium selection criteria to design policies and to determine institutions that are more or less conducive to cooperation \citep[see][]{dalbo2018metainfinite}.
Extending the tools in this literature to games with communication possibilities might improve the results in this literature.
Second, we use word embeddings and machine learning to evaluate the communication content in the laboratory experiments.
There is an emerging literature, to which we add here, using this novel approach to evaluate text corpora in economics \citep[see][]{ash2022wordembeddings}.
Our results provide empirical evidence that this approach fares well in clustering content with similar meaning, even in an experimental context.

In a next step, it would be interesting to identify focal beliefs for different kinds of communication. 
This would allow us to set up equilibrium selection criteria for games with communication possibilities that make point predictions on whether cooperation prevails, such as the well known risk dominance criterion. 
Our results point towards a focal belief of the fraction of three fourths in games with communication.

\bibliographystyle{chicago}
\bibliography{references}

\clearpage
\newpage
\begin{appendix}
\section*{Appendix}\label{sec:appendix}

\section{Theoretical background}\label{sec:model}

The following section presents the theoretical background of the model. Before documenting that the model predicts the cooperation enhancing effect of communication, this section shows that a change in the sucker's payoff $S$ affects the rate of cooperation in games with communication.

\subsection{The effect of a change in the sucker's payoff \textit{S}}\label{sec:model_s}
Equation \ref{eqn:inequalitysuckerspayoff1} ties into Equation \ref{eqn:inequality3} as follows.
\begin{equation}\label{eqn:inequalitysuckerspayoff2}
   \frac{p^{+} \cdot (g-\overline{l})+\overline{l}}{p^{+}\cdot(1+g-\overline{l})+\overline{l}}  > \frac{p^{+} \cdot (g-\underline{l})+\underline{l}}{p^{+}\cdot(1+g-\underline{l})+\underline{l}}
\end{equation}

Equation \ref{eqn:inequalitysuckerspayoff2} results in Equation \ref{eqn:inequalitysuckerspayoff5}.

\begin{equation}\label{eqn:inequalitysuckerspayoff5}
\begin{gathered}
   (p^{+})^2 \cdot g + (p^{+})^2 \cdot g^2 - (p^{+})^2 \cdot \underline{l} \cdot g + p^{+} \cdot g \cdot \underline{l} \\  
   - (p^{+})^2 \cdot \overline{l} - (p^{+})^2 \cdot g \cdot  \overline{l} +  (p^{+})^2 \cdot \overline{l} \cdot  \underline{l}  - p^{+} \cdot  \overline{l} \cdot  \underline{l} \\ 
   + p^{+} \cdot \overline{l} + p^{+} \cdot g \cdot \overline{l}  - p^{+} \cdot \overline{l}  \cdot \underline{l}  + \overline{l}  \cdot \underline{l}
   \\  > \\   
   (p^{+})^2 \cdot g + (p^{+})^2 \cdot g^2 - (p^{+})^2 \cdot \overline{l} \cdot g + p^{+} \cdot g \cdot \overline{l} \\  
   - (p^{+})^2 \cdot \underline{l} - (p^{+})^2 \cdot g \cdot  \underline{l} +  (p^{+})^2 \cdot \overline{l} \cdot  \underline{l}  - p^{+} \cdot  \overline{l} \cdot  \underline{l} \\ 
   + p^{+} \cdot \underline{l} + p^{+} \cdot g \cdot \underline{l}  - p^{+} \cdot \overline{l}  \cdot \underline{l}  + \overline{l}  \cdot \underline{l}
   \end{gathered}
\end{equation}

To simplify Equation \ref{eqn:inequalitysuckerspayoff5}, we subtract $(p^{+})^2 \cdot g$, $(p^{+})^2 \cdot g^2$, $(p^{+})^2 \cdot \overline{l} \cdot  \underline{l}$, $p^{+} \cdot \overline{l} \cdot  g$, $\overline{l} \cdot  \underline{l}$ and $p^{+} \cdot g \cdot  \underline{l}$. 
Further, we add $(p^{+})^2 \cdot \overline{l} \cdot  g$, $(p^{+})^2 \cdot g\cdot  \underline{l}$ and $2 \cdot p^{+} \cdot \overline{l} \cdot  \underline{l}$. 
The result is shown in Equation \ref{eqn:inequalitysuckerspayoff6}.
\begin{equation}\label{eqn:inequalitysuckerspayoff6}
   - (p^{+})^2 \cdot \overline{l} + p^{+} \cdot \overline{l}  
   > 
   - (p^{+})^2 \cdot \underline{l} + p^{+} \cdot \underline{l} 
\end{equation}
Next, to simplify Equation \ref{eqn:inequalitysuckerspayoff6}, we subtract $p^{+}\cdot \underline{l}$ and add $(p^{+})^2 \cdot \overline{l}$. The result is shown in Equation \ref{eqn:inequalitysuckerspayoff7}.
\begin{equation}\label{eqn:inequalitysuckerspayoff7}
   p^{+} \cdot \overline{l}  - p^{+} \cdot \underline{l}
   > 
   (p^{+})^2 \cdot \overline{l} - (p^{+})^2 \cdot \underline{l} 
\end{equation}

Equation \ref{eqn:inequalitysuckerspayoff7} results in Equation \ref{eqn:inequalitysuckerspayoff8}.
\begin{equation}\label{eqn:inequalitysuckerspayoff8}
   p^{+} \cdot ( \overline{l}  -\underline{l} )
   > 
   (p^{+})^2 \cdot ( \overline{l}  -\underline{l} )
\end{equation}

As $( \overline{l}  -\underline{l} )>0$, the division of Equation \ref{eqn:inequalitysuckerspayoff8} by $( \overline{l}  -\underline{l} )$ results in Equation \ref{eqn:inequalitysuckerspayoff9}.
\begin{equation}\label{eqn:inequalitysuckerspayoff9}
   p^{+}
   > 
   (p^{+})^2 
\end{equation}

In a final step, we divide Equation \ref{eqn:inequalitysuckerspayoff9} by $p^{+}$. The result is shown in Equation \ref{eqn:inequalitysuckerspayoff10}.
\begin{equation}\label{eqn:inequalitysuckerspayoff10}
   1
   > 
   p^{+}
\end{equation}

Equation \ref{eqn:inequalitysuckerspayoff10} documents the model predicting that a change in the loss $l$ affects the rate of cooperation ($    \delta^{+}_{\overline{l}} > \delta^{+}_{\underline{l}} $) if and only if $1>p^{+}$. This holds given our assumption.

\subsection{The effect of communication}\label{sec:model_c}

This section documents that the model implies the cooperation enhancing effect of communication. 
Equation \ref{eqn:inequalitycooperation1} ($\delta^{rd} > \delta^{+}$) written out, is shown in Equation \ref{eqn:inequalitycooperation2}.
\begin{equation}\label{eqn:inequalitycooperation2}
   \frac{g+l}{1+g+l} > \frac{p^{+} \cdot (g-l)+l}{p^{+}\cdot(1+g-l)+l}
\end{equation}

Equation \ref{eqn:inequalitycooperation2} results in Equation \ref{eqn:inequalitycooperation6}.
\begin{equation}\label{eqn:inequalitycooperation6}
\begin{gathered}
   g \cdot p^{+} + g^2 \cdot p^{+}  + g\cdot l + l \cdot p^{+}  - l^2 \cdot p^{+} + l^2 \\  > \\   p^{+} \cdot g-p^{+} \cdot l+l + g^2 \cdot p^{+} + g \cdot l  -p^{+} \cdot l^2+l^2
   \end{gathered}
\end{equation}
To simplify Equation \ref{eqn:inequalitycooperation6}, we subtract $l^2$, $g\cdot p^{+}$, $g^2 \cdot p^{+}$ and $l \cdot g$. Further, we add $p^{+} \cdot l^2$. The result is shown in Equation \ref{eqn:inequalitycooperation7}.
\begin{equation}\label{eqn:inequalitycooperation7}
   2 \cdot l \cdot p^{+} > l
\end{equation}
In a final step, we divide Equation \ref{eqn:inequalitycooperation7} by $2 \cdot l$. The result is shown in Equation \ref{eqn:inequalitycooperation8}.
\begin{equation}\label{eqn:inequalitycooperation8}
   p^{+} > \frac{1}{2}
\end{equation}
Equation \ref{eqn:inequalitycooperation8} documents the model predicting the cooperation enhancing effect of communication ($\delta^{rd} > \delta^{+}$) if and only if $p^{+}>\frac{1}{2}$. This holds given our assumption.

\newpage

\section{Instructions}\label{sec:instructions}

In the following, we present our instructions for participants in the \textsc{Comm0} treatment. Parts that appear only in the instructions of a particular treatment are clearly marked as such. Text in \textit{italics} only appears in the instructions if people can communicate. Numbers in square brackets appear in the instructions inherent a sucker's payoff $S$ of $0$ as $0$ and of $70$ as $70$. The (original) instructions for the participants were in German.

\newpage

\begin{singlespace}
\begin{mdframed}[
        linewidth=2pt,%
        frametitle={Instructions},
        outerlinewidth=1.25pt,
        frametitlealignment=\centering
    ]
\end{mdframed}

\noindent Today you are taking part in a decision-making experiment. If you read the following explanations carefully, you can earn money. The amount you receive depends on your decisions and the decisions of other participants.\\
\\
You are not allowed to communicate with other participants for the entire duration of the experiment. We therefore ask you not to talk to each other. Violation of this rule will result in exclusion from the experiment and payment.\\
\\
If there is anything you do not understand, please refer to these experiment instructions again or give us a hand signal. We will then come to you and answer your question personally.\\
\\
During the experiment we do not talk about euros, but about points. The number of points you score during the experiment will be converted into euros as follows:

\begin{center}
    \textbf{180 Points = 1 Euro}
\end{center}

\noindent
At the end of today's experiment, you will receive the points you have achieved from the experiment converted into euros plus 5 euros in \textbf{cash} as basic equipment.\\
\\
The instructions are the same for all participants. On the following pages, we will explain the exact procedure of the experiment.\\
\\
\textbf{The Experiment} \\
\\ 
The experiment consists of 5 independent sub-experiments. All sub-experiments have the same structure. Each sub-experiment consists of several rounds. All rounds have the same structure. In each round you have the opportunity to choose an action.\\
\\
For each sub-experiment, the participants are randomly assigned into groups of 2 persons each. The grouping remains the same within a sub-experiment. Neither you nor the other people learn anything about the identity of the participants in the groups - neither before nor after the experiment.\\
\\
The grouping changes after each sub-experiment. It is ensured that you will meet the same person in a group at most once during the entire experiment. In each new sub-experiment, you will meet a new person whom you have not met before or will meet after the experiment.\\
\\
\textit{At the beginning of each sub-experiment, you can communicate with the other person in your group in writing via chat on the computer. The duration of the chat is limited to 60 seconds. You can write whatever you like in the chat, with the only restriction that you may not give any hint of your identity.}\\ 
\\
\newpage
\noindent\textbf{The Sub-Experiment} \\
\\
A sub-experiment consists of several rounds.\\ 
\\
Exactly how many rounds there are depends on chance. Before each round, a number between 0 and 100 is drawn. For technical reasons, the computer does this. Each number has the same probability of being drawn. If the number is less than 75, a new round starts, if the number is greater than or equal to 75, the sub-experiment is finished and if necessary a new sub-experiment with a new person follows. The random numbers are drawn independently of each other.\\
\\
In each round you can earn points depending on your decision and the decision of the other person you interact with. The points earned in each round are added up and paid out in \textbf{cash} at the end of today's experiment.\\
\\
\textbf{The Round} \\
\\
In each round, you and the other person are simultaneously asked to decide between the two actions \textbf{A} and \textbf{B}. You make your decision for the action \textbf{A} or \textbf{B} by clicking the corresponding red button on the screen with the mouse. After clicking, the decision is irrevocably made. You should make your decision within 30 seconds if possible. After that you will be warned by a flashing display.\\
\\
Your payoff in the round depends on your action and the action of the other person you interact with. The payoff's are as follows:
\begin{table}[H]
\resizebox{\textwidth}{!}
{\begin{tabular}{|c|cc|}
\hline
My decision & \multicolumn{2}{c|}{Payoff if the other person chooses}                                  \\ \cline{2-3} 
                   & \multicolumn{1}{c|}{A}                         & B                          \\ \hline
\multirow{2}{*}{A} & \multicolumn{1}{c|}{For me \textbf{90 Points}}              & For me \textbf{[0] Points}               \\
                   & \multicolumn{1}{c|}{For the other person \textbf{90 Points}} & For the other person \textbf{100 Points} \\ \hline
\multirow{2}{*}{B} & \multicolumn{1}{c|}{For me \textbf{100 Points}}             & For me \textbf{80 Points}             \\
                   & \multicolumn{1}{c|}{For the other person \textbf{[0] Points}} & For the other person \textbf{80 Points}  \\ \hline
\end{tabular}}
\end{table}

\noindent At the end of each round, you will learn the decision of the other person you are interacting with, your scored points in that round, and the sum of the scored points in the current sub-experiment. The information remains visible for 30 seconds. However, you can exit the information screen before that by clicking the gray OK button. When all participants have left the screen by clicking the gray OK button, but after 30 seconds at the latest, the next round will begin, if applicable.\\
\\
We will ask you to answer some comprehension questions on the computer in a moment. This is to make sure that all participants have understood this instructions well.\\
\\
After the experiment, we will ask you to fill out a short questionnaire on the computer. After that you will receive your payout.\\

\end{singlespace}

\newpage

\section{Quiz}\label{sec:quiz}

In the following, we present our comprehension questions. \Paste{quiz} The comprehension questions are:

\noindent(1) How many people (including you) are in a group?

\noindent(2) What payoff do you get in a round if you and the other person choose "B"? 

\noindent(3) What payoff do you get in a round if you and the other person choose "A"? 

\noindent(4) If the random number is less than what number, a new round starts?

\noindent(5) The random numbers are drawn independently?

\noindent(6) What is the total payoff you would get in a sub-experiment if the sub-experiment lasted 4 rounds and you and the other person always chose "A"?

\noindent(7) In each new sub-experiment, you will meet a new person whom you have not met before or will meet after the experiment?

\noindent(8) What payoff do you get in a round if you choose "A" and the other person chooses "B"?

\newpage

\section{Data}\label{sec:additionalresults}

In the following, we present the data discussed in this article. 

\newpage

\subsection{The total-within clusters sum of square}
\begin{figure}[H]
	\centering
	\includegraphics[width=0.7\textwidth]{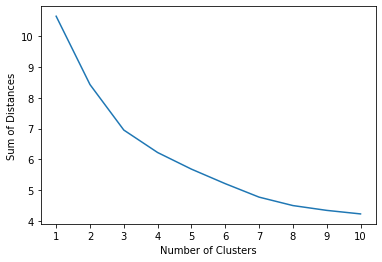}
	\caption{The \Paste{elbow}}
	\label{fig:elbowcurve}
\end{figure}

A need for any $k$-means algorithm lies in choosing the number of cluster $k$. To choose the number of cluster $k$, we rely on the total-within clusters sum of square. Figure \ref{fig:elbowcurve} shows the \Copy{elbow}{total-within clusters sum of square per number of cluster $k$.} It is apparent from this figure that an increasing number of cluster $k$ led to a lower total-within clusters sum of square. The statistically optimal number of cluster lies at the point where choosing one more cluster does not reduce the the total-within clusters sum of square much further. In Figure \ref{fig:elbowcurve}, this is the number of cluster $k$ where the elbow lies. It is apparent from this figure, that the number of cluster $k$ that we consider for the $k$-means algorithm is $3$. Two of those cluster are meaningful: they consider a conversation. In the third cluster, both subjects only greet each other without further intention.\footnote{The human-coders, as described in the main text, find a similar third cluster that contains greetings between the participants only.} Thus, for the analysis of the communication content, we just consider the two meaningful cluster.

\newpage

\subsection{Belief split up by treatment}
\begin{table}[H]
\begin{center}
\footnotesize
\subfloat[First supergame\label{table:belieffirstsupergame}]{
\begin{tabular}{cccccc}
\hline
                                     &                              & \multicolumn{4}{c}{\multirow{2}{*}{Treatment}}                                                                                                                                                                                                                                                            \\
                                     &                              & \multicolumn{4}{c}{}                                                                                                                                                                                                                                                                                      \\ \cline{3-6} 
                                     &                              & \multirow{2}{*}{NoComm70}                                                & \multirow{2}{*}{NoComm0}                                                 & \multirow{2}{*}{Comm70}                                                  & \multirow{2}{*}{Comm0}                                                   \\
                                     &                              &                                                                          &                                                                          &                                                                          &                                                                          \\ \hline
\multirow{5}{*}{Belief}              & \multirow{3}{*}{Mean}        & \multirow{3}{*}{\begin{tabular}[c]{@{}c@{}}54.59\\ (31.16)\end{tabular}} & \multirow{3}{*}{\begin{tabular}[c]{@{}c@{}}35.42\\ (29.37)\end{tabular}} & \multirow{3}{*}{\begin{tabular}[c]{@{}c@{}}81.51\\ (23.83)\end{tabular}} & \multirow{3}{*}{\begin{tabular}[c]{@{}c@{}}84.99\\ (17.45)\end{tabular}} \\
                                     &                              &                                                                          &                                                                          &                                                                          &                                                                          \\
                                     &                              &                                                                          &                                                                          &                                                                          &                                                                          \\
                                     & \multirow{2}{*}{Median}      & \multirow{2}{*}{64.69}                                                   & \multirow{2}{*}{30.47}                                                   & \multirow{2}{*}{89.94}                                                   & \multirow{2}{*}{88.66}                                                   \\
                                     &                              &                                                                          &                                                                          &                                                                          &                                                                          \\ \hline
\multirow{4}{*}{\textgreater Median} & \multirow{2}{*}{Cooperation} & \multirow{2}{*}{14}                                                      & \multirow{2}{*}{14}                                                      & \multirow{2}{*}{15}                                                      & \multirow{2}{*}{15}                                                      \\
                                     &                              &                                                                          &                                                                          &                                                                          &                                                                          \\
                                     & \multirow{2}{*}{Defection}   & \multirow{2}{*}{1}                                                       & \multirow{2}{*}{7}                                                       & \multirow{2}{*}{0}                                                       & \multirow{2}{*}{0}                                                       \\
                                     &                              &                                                                          &                                                                          &                                                                          &                                                                          \\ \hline
\multirow{4}{*}{\textless Median}    & \multirow{2}{*}{Cooperation} & \multirow{2}{*}{2}                                                       & \multirow{2}{*}{1}                                                       & \multirow{2}{*}{12}                                                      & \multirow{2}{*}{11}                                                      \\
                                     &                              &                                                                          &                                                                          &                                                                          &                                                                          \\
                                     & \multirow{2}{*}{Defection}   & \multirow{2}{*}{13}                                                       & \multirow{2}{*}{20}                                                      & \multirow{2}{*}{3}                                                       & \multirow{2}{*}{4}                                                       \\
                                     &                              &                                                                          &                                                                          &                                                                          &                                                                          \\ \hline
\end{tabular}
}
	\vspace*{0.25in}

\subfloat[Final supergame\label{table:belieffinalsupergame}]{
\begin{tabular}{cccccc}
\hline
                                     &                              & \multicolumn{4}{c}{\multirow{2}{*}{Treatment}}                                                                                                                                                                                                                                                            \\
                                     &                              & \multicolumn{4}{c}{}                                                                                                                                                                                                                                                                                      \\ \cline{3-6} 
                                     &                              & \multirow{2}{*}{NoComm70}                                                & \multirow{2}{*}{NoComm0}                                                 & \multirow{2}{*}{Comm70}                                                  & \multirow{2}{*}{Comm0}                                                   \\
                                     &                              &                                                                          &                                                                          &                                                                          &                                                                          \\ \hline
\multirow{5}{*}{Belief}              & \multirow{3}{*}{Mean}        & \multirow{3}{*}{\begin{tabular}[c]{@{}c@{}}40.56\\ (28.08)\end{tabular}} & \multirow{3}{*}{\begin{tabular}[c]{@{}c@{}}15.88\\ (25.51)\end{tabular}} & \multirow{3}{*}{\begin{tabular}[c]{@{}c@{}}81.80\\ (23.25)\end{tabular}} & \multirow{3}{*}{\begin{tabular}[c]{@{}c@{}}68.33\\ (37.76)\end{tabular}} \\
                                     &                              &                                                                          &                                                                          &                                                                          &                                                                          \\
                                     &                              &                                                                          &                                                                          &                                                                          &                                                                          \\
                                     & \multirow{2}{*}{Median}      & \multirow{2}{*}{43.30}                                                   & \multirow{2}{*}{0.73}                                                    & \multirow{2}{*}{90.99}                                                   & \multirow{2}{*}{84.75}                                                   \\
                                     &                              &                                                                          &                                                                          &                                                                          &                                                                          \\ \hline
\multirow{4}{*}{\textgreater Median} & \multirow{2}{*}{Cooperation} & \multirow{2}{*}{13}                                                      & \multirow{2}{*}{8}                                                       & \multirow{2}{*}{15}                                                      & \multirow{2}{*}{13}                                                      \\
                                     &                              &                                                                          &                                                                          &                                                                          &                                                                          \\
                                     & \multirow{2}{*}{Defection}   & \multirow{2}{*}{2}                                                       & \multirow{2}{*}{13}                                                      & \multirow{2}{*}{0}                                                       & \multirow{2}{*}{2}                                                       \\
                                     &                              &                                                                          &                                                                          &                                                                          &                                                                          \\ \hline
\multirow{4}{*}{\textless Median}    & \multirow{2}{*}{Cooperation} & \multirow{2}{*}{3}                                                       & \multirow{2}{*}{0}                                                       & \multirow{2}{*}{15}                                                      & \multirow{2}{*}{8}                                                       \\
                                     &                              &                                                                          &                                                                          &                                                                          &                                                                          \\
                                     & \multirow{2}{*}{Defection}   & \multirow{2}{*}{12}                                                      & \multirow{2}{*}{21}                                                      & \multirow{2}{*}{0}                                                       & \multirow{2}{*}{7}                                                       \\
                                     &                              &                                                                          &                                                                          &                                                                          &                                                                          \\ \hline
\end{tabular}
}
\caption[Mean and median belief as well as above- and below-median cooperation count split up by treatment in the first and final supergame.]{Mean and median belief as well as above- and below-median cooperation count split up by treatment in the first and final supergame. Numbers in brackets are standard deviations.}
\label{table:beliefmedian}
\end{center}
\end{table}

\newpage

\subsection{Cooperation per supergame split up by treatment}

Table \ref{table:cooperationsupergames}: the $p$-value between \textsc{NoComm70} and \textsc{NoComm0} in all rounds is $0.012$, $0.036$, $0.017$, $0.024$, $0.019$ and $0.007$ \Copy{firstusw}{in the first, second, third, fourth, fith and over all supergames, respectively.}  The $p$-value between \textsc{Comm70} and \textsc{Comm0} in all rounds is $0.375$, $0.090$, $0.001$, $0.017$, $0.193$ and $0.072$ \Paste{firstusw} The $p$-value between \textsc{Comm70} and \textsc{NoComm70} in all rounds is $0.005$, $0.004$, $0.005$, $0.010$, $0.018$ and $0.006$ \Paste{firstusw} The $p$-value between \textsc{Comm0} and \textsc{NoComm0} in all rounds is $0.003$, $0.003$, $0.005$, $0.009$, $0.010$ and $0.005$ \Paste{firstusw}

\newpage

\begin{landscape}
\begin{table}[H]
\begin{center}
\footnotesize
\subfloat[First round\label{table:cooperationsupergamesfirstrounds}]{
\begin{tabular}{ccccc}
\hline
\multirow{2}{*}{Supergame} & \multicolumn{4}{c}{\multirow{2}{*}{Treatment}}                                                                                                                                                                                                                                                      \\
                           & \multicolumn{4}{c}{}                                                                                                                                                                                                                                                                                \\ \cline{2-5} 
                           & \multirow{2}{*}{NoComm70}                                               & \multirow{2}{*}{NoComm0}                                                & \multirow{2}{*}{Comm70}                                                & \multirow{2}{*}{Comm0}                                                 \\
                           &                                                                         &                                                                         &                                                                        &                                                                        \\ \hline
\multirow{3}{*}{1}         & \multirow{3}{*}{\begin{tabular}[c]{@{}c@{}}0.53\\ (0.08)\end{tabular}}  & \multirow{3}{*}{\begin{tabular}[c]{@{}c@{}}0.36\\ (0.24)\end{tabular}}  & \multirow{3}{*}{\begin{tabular}[c]{@{}c@{}}0.90\\ (0.09)\end{tabular}} & \multirow{3}{*}{\begin{tabular}[c]{@{}c@{}}0.87\\ (0.14)\end{tabular}} \\
                           &                                                                         &                                                                         &                                                                        &                                                                        \\
                           &                                                                         &                                                                         &                                                                        &                                                                        \\
\multirow{3}{*}{2}         & \multirow{3}{*}{\begin{tabular}[c]{@{}c@{}}0.43\\ (0.15)\end{tabular}}  & \multirow{3}{*}{\begin{tabular}[c]{@{}c@{}}0.21\\ (0.27)\end{tabular}}  & \multirow{3}{*}{\begin{tabular}[c]{@{}c@{}}1.00\\ (0.00)\end{tabular}} & \multirow{3}{*}{\begin{tabular}[c]{@{}c@{}}0.83\\ (0.24)\end{tabular}} \\
                           &                                                                         &                                                                         &                                                                        &                                                                        \\
                           &                                                                         &                                                                         &                                                                        &                                                                        \\
\multirow{3}{*}{3}         & \multirow{3}{*}{\begin{tabular}[c]{@{}c@{}}0.50\\ (0.26)\end{tabular}}  & \multirow{3}{*}{\begin{tabular}[c]{@{}c@{}}0.19\\ (0.20)\end{tabular}}  & \multirow{3}{*}{\begin{tabular}[c]{@{}c@{}}1.00\\ (0.00)\end{tabular}} & \multirow{3}{*}{\begin{tabular}[c]{@{}c@{}}0.73\\ (0.19)\end{tabular}} \\
                           &                                                                         &                                                                         &                                                                        &                                                                        \\
                           &                                                                         &                                                                         &                                                                        &                                                                        \\
\multirow{3}{*}{4}         & \multirow{3}{*}{\begin{tabular}[c]{@{}c@{}}0.53\\ (0.27)\end{tabular}}  & \multirow{3}{*}{\begin{tabular}[c]{@{}c@{}}0.17\\ (0.22)\end{tabular}}  & \multirow{3}{*}{\begin{tabular}[c]{@{}c@{}}0.93\\ (0.15)\end{tabular}} & \multirow{3}{*}{\begin{tabular}[c]{@{}c@{}}0.77\\ (0.28)\end{tabular}} \\
                           &                                                                         &                                                                         &                                                                        &                                                                        \\
                           &                                                                         &                                                                         &                                                                        &                                                                        \\
\multirow{3}{*}{5}         & \multirow{3}{*}{\begin{tabular}[c]{@{}c@{}}0.53\\ (0.27)\end{tabular}}  & \multirow{3}{*}{\begin{tabular}[c]{@{}c@{}}0.19\\ (0.26)\end{tabular}}  & \multirow{3}{*}{\begin{tabular}[c]{@{}c@{}}1.00\\ (0.00)\end{tabular}} & \multirow{3}{*}{\begin{tabular}[c]{@{}c@{}}0.70\\ (0.34)\end{tabular}} \\
                           &                                                                         &                                                                         &                                                                        &                                                                        \\
                           &                                                                         &                                                                         &                                                                        &                                                                        \\ \hline
\multirow{3}{*}{Mean}      & \multirow{3}{*}{\begin{tabular}[c]{@{}c@{}}0.51 \\ (0.17)\end{tabular}} & \multirow{3}{*}{\begin{tabular}[c]{@{}c@{}}0.22 \\ (0.23)\end{tabular}} & \multirow{3}{*}{\begin{tabular}[c]{@{}c@{}}0.97\\ (0.04)\end{tabular}} & \multirow{3}{*}{\begin{tabular}[c]{@{}c@{}}0.78\\ (0.22)\end{tabular}} \\
                           &                                                                         &                                                                         &                                                                        &                                                                        \\
                           &                                                                         &                                                                         &                                                                        &                                                                        \\ \hline
\end{tabular}
}
\vspace*{0.25in}
\subfloat[All rounds\label{table:cooperationsupergamesallrounds}]{
\begin{tabular}{ccccc}
\hline
\multirow{2}{*}{Supergame} & \multicolumn{4}{c}{\multirow{2}{*}{Treatment}}                                                                                                                                                                                                                                                       \\
                           & \multicolumn{4}{c}{}                                                                                                                                                                                                                                                                                 \\ \cline{2-5} 
                           & \multirow{2}{*}{NoComm70}                                               & \multirow{2}{*}{NoComm0}                                                & \multirow{2}{*}{Comm70}                                                 & \multirow{2}{*}{Comm0}                                                 \\
                           &                                                                         &                                                                         &                                                                         &                                                                        \\ \hline
\multirow{3}{*}{1}         & \multirow{3}{*}{\begin{tabular}[c]{@{}c@{}}0.52\\ (0.04)\end{tabular}}  & \multirow{3}{*}{\begin{tabular}[c]{@{}c@{}}0.25\\ (0.23)\end{tabular}}  & \multirow{3}{*}{\begin{tabular}[c]{@{}c@{}}0.80\\ (0.60)\end{tabular}}  & \multirow{3}{*}{\begin{tabular}[c]{@{}c@{}}0.77\\ (0.22)\end{tabular}} \\
                           &                                                                         &                                                                         &                                                                         &                                                                        \\
                           &                                                                         &                                                                         &                                                                         &                                                                        \\
\multirow{3}{*}{2}         & \multirow{3}{*}{\begin{tabular}[c]{@{}c@{}}0.39\\ (0.12)\end{tabular}}  & \multirow{3}{*}{\begin{tabular}[c]{@{}c@{}}0.15\\ (0.18)\end{tabular}}  & \multirow{3}{*}{\begin{tabular}[c]{@{}c@{}}1.00\\ (0.00)\end{tabular}}  & \multirow{3}{*}{\begin{tabular}[c]{@{}c@{}}0.82\\ (0.27)\end{tabular}} \\
                           &                                                                         &                                                                         &                                                                         &                                                                        \\
                           &                                                                         &                                                                         &                                                                         &                                                                        \\
\multirow{3}{*}{3}         & \multirow{3}{*}{\begin{tabular}[c]{@{}c@{}}0.46\\ (0.20)\end{tabular}}  & \multirow{3}{*}{\begin{tabular}[c]{@{}c@{}}0.12\\ (0.16)\end{tabular}}  & \multirow{3}{*}{\begin{tabular}[c]{@{}c@{}}0.94\\ (0.09)\end{tabular}}  & \multirow{3}{*}{\begin{tabular}[c]{@{}c@{}}0.67\\ (0.17)\end{tabular}} \\
                           &                                                                         &                                                                         &                                                                         &                                                                        \\
                           &                                                                         &                                                                         &                                                                         &                                                                        \\
\multirow{3}{*}{4}         & \multirow{3}{*}{\begin{tabular}[c]{@{}c@{}}0.43\\ (0.33)\end{tabular}}  & \multirow{3}{*}{\begin{tabular}[c]{@{}c@{}}0.11\\ (0.16)\end{tabular}}  & \multirow{3}{*}{\begin{tabular}[c]{@{}c@{}}0.95\\ (0.08)\end{tabular}}  & \multirow{3}{*}{\begin{tabular}[c]{@{}c@{}}0.65\\ (0.35)\end{tabular}} \\
                           &                                                                         &                                                                         &                                                                         &                                                                        \\
                           &                                                                         &                                                                         &                                                                         &                                                                        \\
\multirow{3}{*}{5}         & \multirow{3}{*}{\begin{tabular}[c]{@{}c@{}}0.39\\ (0.30)\end{tabular}}  & \multirow{3}{*}{\begin{tabular}[c]{@{}c@{}}0.10\\ (0.13)\end{tabular}}  & \multirow{3}{*}{\begin{tabular}[c]{@{}c@{}}0.89 \\ (0.13)\end{tabular}} & \multirow{3}{*}{\begin{tabular}[c]{@{}c@{}}0.67\\ (0.40)\end{tabular}} \\
                           &                                                                         &                                                                         &                                                                         &                                                                        \\
                           &                                                                         &                                                                         &                                                                         &                                                                        \\ \hline
\multirow{3}{*}{Mean}       & \multirow{3}{*}{\begin{tabular}[c]{@{}c@{}}0.44 \\ (0.18)\end{tabular}} & \multirow{3}{*}{\begin{tabular}[c]{@{}c@{}}0.15 \\ (0.16)\end{tabular}} & \multirow{3}{*}{\begin{tabular}[c]{@{}c@{}}0.92\\ (0.04)\end{tabular}}  & \multirow{3}{*}{\begin{tabular}[c]{@{}c@{}}0.72\\ (0.26)\end{tabular}} \\
                           &                                                                         &                                                                         &                                                                         &                                                                        \\
                           &                                                                         &                                                                         &                                                                         &                                                                        \\ \hline
\end{tabular}
}
\caption[The rate of cooperation in the first round and all rounds over supergames split up by treatments.]{The rate of cooperation in the first round and all rounds over supergames split up by treatments. Numbers in brackets are standard deviations.}
\label{table:cooperationsupergames}
\end{center}
\end{table}
\end{landscape}

\newpage

\section{Original German tokens in their corresponding Figure}\label{sec:figures_in_german_kmeans}
\parindent0pt
\parskip6pt

Here, we show the original German tokens in their corresponding figure. We translated the tokens only after the analyzes of the chats.

\begin{figure}[H]
	\centering
	\includegraphics[width=1\linewidth]{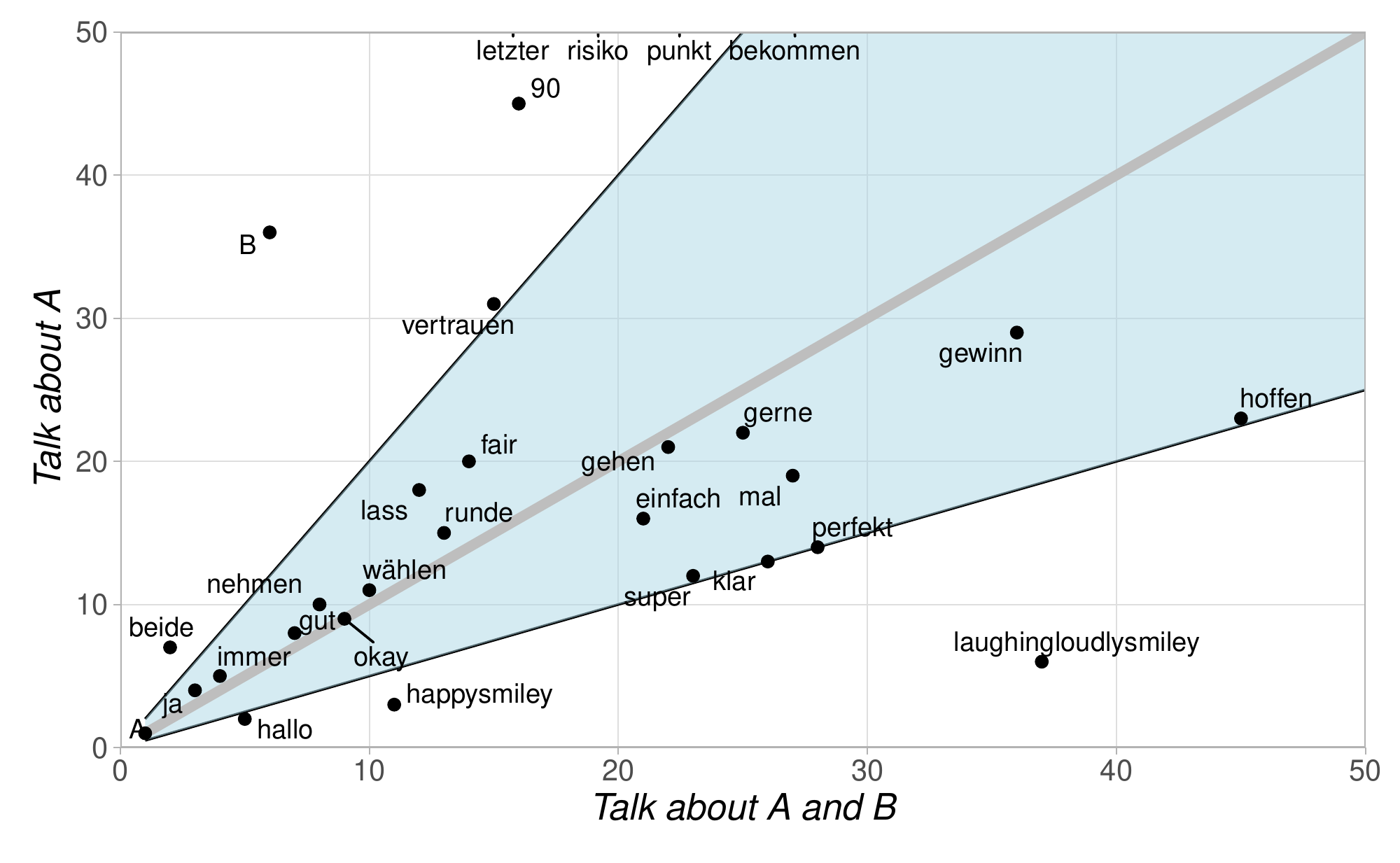}
\caption{Frequency rankings of the 30 most used German tokens in both clusters.}\label{fig:figures_in_german_kmeans}
\end{figure}

\end{appendix}
\end{document}